\documentclass[aps,prx,reprint,superscriptaddress]{revtex4-1}

\usepackage{graphicx}
\usepackage{amsmath}
\usepackage{amsfonts}
\usepackage{amssymb}
\usepackage{bbm}
\usepackage[]{units}
\usepackage{hyperref}
\usepackage{braket}
\usepackage{xcolor}

\newcommand{\nnb}{\nonumber \\}

\newcommand{\bv}{\left( \begin{array}{c}}
\newcommand{\ev}{\end{array} \right)}
\newcommand{\E}{\mathrm{e}}

\newcommand{\st}[1]{_{\text{#1}}}

\newcommand{\I}{\mathrm{i}}

\begin{document}
\title{A quadrupolar exchange-only spin qubit}
\author{Maximilian Russ}
\affiliation{Department of Physics, University of Konstanz, D-78457 Konstanz, Germany}
\author{J. R. Petta}
\affiliation{Department of Physics, Princeton University, Princeton, New Jersey 08544, USA}
\author{Guido Burkard}
\affiliation{Department of Physics, University of Konstanz, D-78457 Konstanz, Germany}
\begin{abstract}
We propose a quadrupolar exchange-only spin (QUEX) qubit that is highly
robust against charge noise and nuclear spin dephasing, the dominant decoherence mechanisms in quantum dots. The qubit consists of four electrons trapped in three quantum dots, and operates in a decoherence-free subspace to mitigate dephasing due to nuclear spins. To reduce sensitivity to charge noise, the qubit can be completely operated at an extended charge noise sweet spot that is first-order insensitive to electrical fluctuations. Due to on-site exchange mediated by the Coulomb interaction, the qubit energy splitting is electrically controllable and can amount to several GHz even in the ``off" configuration, making it compatible with conventional microwave cavities.
\end{abstract}
\maketitle
%%%%%%%%%%%%%%%%%%%%%%%%%%%%%%%%%%%%%%%%
\paragraph*{Introduction}
Electron spin qubits in semiconductor quantum dots have recently demonstrated their capability as components in a working quantum processor~\cite{Veldhorst2015,Yoneda2017,Zajac2017,Watson2018}. With simple quantum algorithms having now been demonstrated, there is a strong motivation for building a large-scale quantum computer using spin qubits due to their long intrinsic coherence times and fast gate operation times~\cite{Loss1998,Hanson2007,Kloeffel2013,Zwanenburg2013}. Spin qubits in silicon additionally benefit from state-of-the-art industrial nanofabrication techniques for scalability and the possibility of isotopic enrichment to increase coherence times. While there are many different implementations of spin qubits, the exchange-only qubit~\cite{Nature2000,Laird2010,Gaudreau2012,Medford2013,Medford2013b,Eng2015,Landig2017} is unique since it can be fully controlled using dc gate voltage pulses. The decoherence-free subspace encoding also makes exchange-only spin qubits insensitive to overall (long-wavelength) magnetic field fluctuations \cite{Nature2000,Bacon2000,Kempe2001}. However, all experimental demonstrations until now suffer from decoherence due to charge noise and local (short-wavelength) magnetic field gradient (LMFG) noise, thus limiting the performance of the qubit~\cite{Medford2013,Medford2013b,Eng2015,Malinowski2017,Landig2017}. 

Protection against charge noise is provided by operating qubits at a so-called charge noise ``sweet spot'', a point of operation which is first-order insensitive to electric fluctuations~\cite{Taylor2013,Fei2015,Russ2015,Russ2016,Shim2016,Reed2016,Martins2016,Russ2017,Malinowski2017}. However, the energy splitting at these sweet spots is too small to couple the qubit to conventional superconducting resonators required for long-distance entanglement protocols~\cite{Imamoglu1999,Childs2000,Childress2004,Burkard2006,Wallraff2007,Leek2009,Russ2015b,Srinivasa2016,Russ2016}. Moreover, three-spin exchange-only qubits are sensitive to LMFG noise \cite{Hung2014,Mehl2015,Russ2017,Peterfalvi2017}, which can arise from fluctuating nuclear spins or Meissner expulsion of magnetic fields near superconducting gates. These gradients limit spin coherence times and can result in leakage, i.e., the loss of information into the non-computational subspace. Additionally, while a single exchange-only qubit or hybrid qubit is insensitive to fluctuations in the overall magnetic field, an array of exchange coupled exchange-only (hybrid) qubits does not benefit from this protection since each exchange-only qubit can acquire a slightly different phase.

\begin{figure}
\begin{center}
\includegraphics[width=1.\columnwidth]{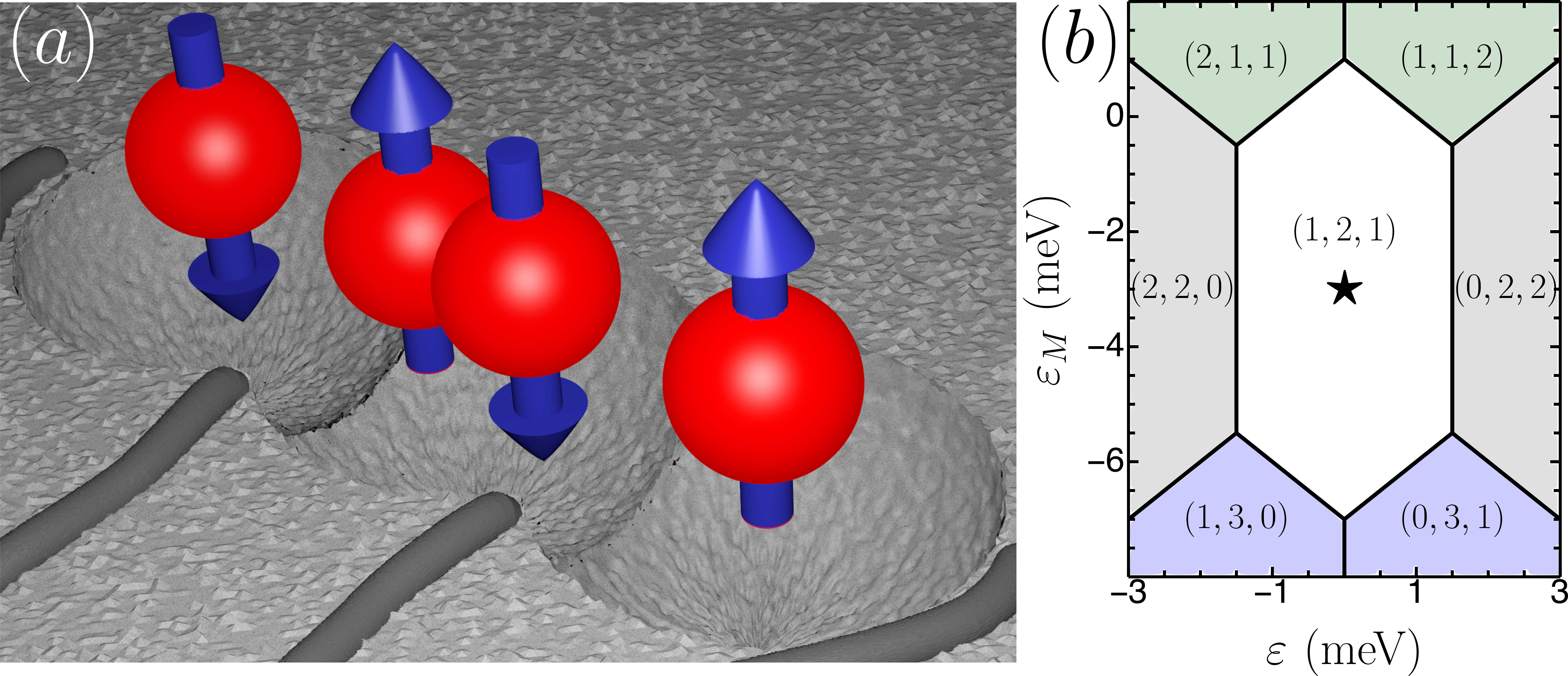}
\caption{(a) Illustration of a four-electron spin qubit residing in an electrostatically defined triple quantum dot (TQD). The four spins are coupled via inter-dot and onsite exchange interaction. The center dot is occupied by two electrons, giving rise to a large and electrostatically tunable energy splitting. (b) TQD charge stability diagram as a function of $\varepsilon$ and $\varepsilon_{M}$. The optimal operating point is marked by the star.}
\label{fig:main1}
\end{center}
\end{figure}

In this Letter, we propose a quadrupolar exchange-only spin (QUEX) qubit that allows for universal quantum computation with high-speed qubit operations and very long coherence times. 
%The QUEX qubit combines the protection against decoherence of the single electron charge quadrupole qubit~\cite{Friesen2017,Kornich2018} with the fast control of the exchange-only qubit~\cite{Nature2000,Laird2010,Gaudreau2012,Medford2013,Taylor2013,Medford2013b,Eng2015,Poulin2015,Fei2015,Russ2015,Shim2016,Russ2016,Russ2017,Malinowski2017} and the hybrid qubit~\cite{Shi2012,Koh2012,Cao2016,Thorgrimsson2016,Russ2017,Wang2017}. 
The QUEX qubit is operated in the 4 electron regime, with the 4 electrons distributed on 3 series-coupled semiconductor quantum dots [see Fig.~\ref{fig:main1}(a)] and is encoded in the low-energy subspace with total spin $S=0$. This encoding makes the QUEX qubit insensitive to overall (long-wavelength) magnetic fields, first-order insensitive to LMFG noise \cite{Bacon2000,Kempe2001,Sala2017}, and no global phases are acquired from long-range magnetic fields, thus there is no desynchronization problem if used in a large-scale array. Compared to the singlet-singlet qubit~\cite{Sala2017}, the QUEX qubit offers a simplified architecture requiring only three dots and two detuning parameters.
As with the exchange-only qubit~\cite{Nature2000}, single-qubit operations in the QUEX qubit can be driven using either dc pulses or ac modulation. 
Compared to the exchange-only~\cite{Nature2000,Laird2010,Gaudreau2012,Medford2013,Taylor2013,Medford2013b,Eng2015,Poulin2015,Fei2015,Russ2015,Shim2016,Russ2016,Russ2017,Malinowski2017} and the hybrid~\cite{Shi2012,Koh2012,Cao2016,Thorgrimsson2016,Russ2017,Wang2017} qubit, the QUEX qubit offers an increased protection against charge noise~\cite{Thorgrimsson2016,Mi2018a} due to an extended charge noise sweet spot. This arises from the addition of the fourth electron which flattens the energy bands and provides a qubit energy splitting at the sweet spot which is electrically controllable and can amount to several GHz even in the ``off" configuration, thus, making it compatible with conventional superconducting resonators. 
Building on recent experiments showing that strong coupling of semiconductor qubits~\cite{Viennot2016,Mi2016b,Stockklauser2017,Mi2018,Landig2017,Samkharadze2017} to electromagnetic resonators is feasible, we present an electrically switchable two-qubit interaction~\cite{Srinivasa2016} making use of the large and strongly tunable qubit splitting.

\begin{figure}
\begin{center}
\includegraphics[width=1\columnwidth]{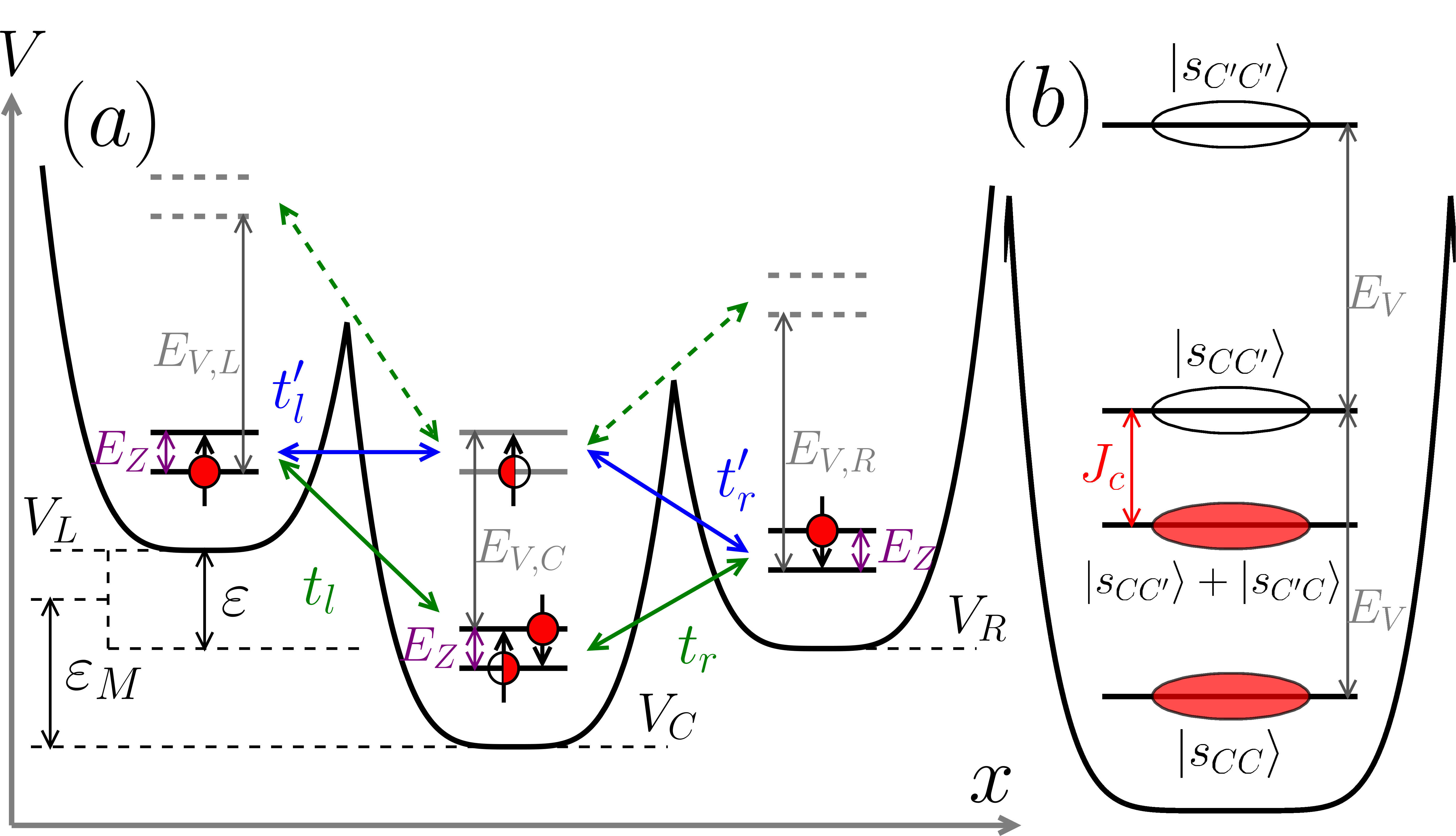}
\caption{(a) TQD confining potential $V(x)$. The left and right dots each contain one electron in the ground state (black level). The middle dot contains two electrons, one of which can occupy an excited valley state (grey level) $E_{V,C}$ above the ground state. The electrons are allowed to hop with valley conserving (green), $t_{l,r}$, and valley non-conserving (blue) tunneling, $t^{\prime}_{l,r}$. Tunneling from the center dot to an excited valley state (grey dashed) in the outer dots is energetically unfavorable.
(b) Energy level diagram of the center quantum dot filled with two electrons for total spin $S_{z}=0$. The lowest energy level is the spin-singlet state $\ket{s_{CC}}$ in the valley ground state. The first excited state $\ket{s_{CC^{\prime}}}+\ket{s_{C^{\prime}C}}$ is a triplet-like state and occupies a valley ground and excited state but its energy is lowered by the electrostatically tunable onsite Coulomb-exchange coupling $J_{c}$.}
\label{fig:main2}
\end{center}
\end{figure}

\paragraph*{Qubit design}
A defining feature of the QUEX qubit is its use of the valley degree of freedom in Si to achieve an energy splitting that is compatible with conventional superconducting cavities with $\unit[4-10]{GHz}$ resonance frequencies. The QUEX qubit is implemented in a TQD that contains a total of four electrons in the (1,2,1) charge configuration [see Figs.~\ref{fig:main1}~and~\ref{fig:main2}(a)]. Here $(N_{L},N_{C},N_{R})$ denote the number of electrons confined in each of the three dots. For later convenience we define the dipolar detuning $\varepsilon\equiv(V_{L}-V_{R})/2$ and the quadrupolar detuning $\varepsilon_{M}\equiv V_{C}-(V_{L}+V_{R})/2$ [see Fig.~\ref{fig:main2}~(a)].

Silicon quantum dots typically have relatively large orbital energies $E\st{orb}$ = $\unit[3-5]{meV}$~\cite{Yang2012,Zajac2016}. In contrast, the smaller valley splittings $E_{V,C}$ = $\unit[20-250]{\mu eV}$ are compatible with microwave frequency photons (1 GHz $\sim$ 4.2 $\mu$eV)~\cite{Yang2012,Mi2017}. In the QUEX qubit, the valley degree of freedom plays a crucial role in the middle dot, as it adds a level that can be treated as an additional fourth pseudo-dot. An external (homogeneous) magnetic field with Zeeman splitting $E_{Z}\ll E_{V,C}\ll E\st{orb}$ energetically separates states according to $S_{z}$. The QUEX qubit resides in the two lowest energy levels in the $S=S_{z}=0$ subspace with the two following logical spin qubit states~\cite{Lidar2012,Sala2017}
\begin{align}
\ket{0}&=\ket{s_{LR}}\ket{s_{CC}}
\label{eq:state0}\\
\ket{1}&=\frac{1}{\sqrt{3}}(\ket{s_{LC^{\prime}}}\ket{s_{CR}}+\ket{s_{LC}}\ket{s_{C^{\prime}R}}).
\label{eq:state1}
\end{align}
Here, $\ket{s_{\mu\nu}}$ denotes the singlet state formed by two electrons in orbitals $\mu$ and $\nu$ with $\mu,\nu=L,C,C^{\prime},R$ where $L$ ($R$) reside in the left QD (right QD) and $C,(C^{\prime})$ reside in the lower (upper) valley in the center QD. Note that both qubit states are constructed using only two-electron singlet states making them resilient to LMFG.

We describe the dynamics of the QUEX qubit by the following effective Hamiltonian
\begin{align}
%H &= J_{0}\ket{0}\bra{0}+(J_{1}+E_{V,C}-J_{C})\ket{1}\bra{1} + J_{x} \ket{1}\bra{0}+ J_{x}^{*} \ket{0}\bra{1}.
H &= J_{0}\ket{0}\bra{0}+(J_{1}+E_{V,C}-J_{C})\ket{1}\bra{1} + J_{x} \sigma_{+}+ J_{x}^{*} \sigma_{-},
\label{eq:Hamqubit}
\end{align}
with the qubit raising and lowering operator $\sigma_{+}=\ket{0}\bra{1}$ and $\sigma_{-}=\sigma_{+}^\dagger=\ket{1}\bra{0}$.
Here, $J_{0}$ and $J_{1}$ are real-valued and $J_{x}$ complex-valued exchange couplings from virtual tunneling processes to states with a (2,2,0), (2,1,1), (1,3,0), (0,3,1), (1,1,2) and (0,2,2) charge configuration, $E_{V,C}$ is the valley splitting in the center QD, and $J_{C}$ is the onsite Coulomb-exchange coupling between the electrons occupying the center QD. The phase of $J_{x}$ depends on the phase an electron acquires by consecutively tunneling from the lower to the excited valley state on the center dot via an intermediate state on the left or right dot (see Fig.~\ref{fig:main2}). For later convenience, we can also write $J_{\mu}=\sum_{\nu}J_{\mu}^{\nu}$ with $\mu=0,1,x$ and $\nu=l,l^{\prime},r,r^{\prime}$ where $J_{\mu}^{l}$ ($J_{\mu}^{L}$) denotes exchange coupling via valley conserving (non-conserving) tunneling of an electron from the center QD to the left QD and similarly for $J_{\mu}^{r}$ ($J_{\mu}^{R}$) to the right QD. The qubit splitting for $|J_{x}|\ll |E_{V,C}-J_{C}|$ is given by
\begin{align}
\omega_{q}\approx (J_{1}+E_{V,C}-J_{C}-J_{0})+\frac{|J_{x}|^{2}}{2(J_{1}+E_{V,C}-J_{C}-J_{0})}.
\end{align}
The explicit expressions for the exchange parameters and a detailed derivation of the effective Hamiltonian~\eqref{eq:Hamqubit} can be found in~\ref{app:der}.

\begin{figure}
\begin{center}
\includegraphics[width=1\columnwidth]{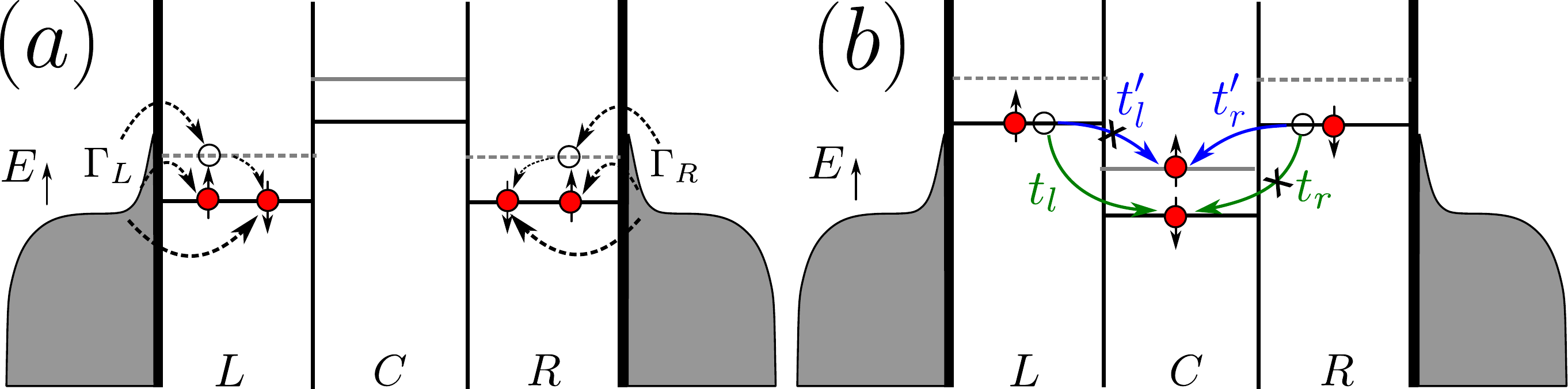}
\caption{Protocol for initializing the QUEX qubit: First the (2,0,2) charge configuration regime (a) is prepared by loading the left and right QD with two electrons in the spin-up and spin-down state of the lowest valley. (b) Adiabatic tuning from the (2,0,2) to the (1,2,1) charge state directly maps $\ket{\text{init}}\rightarrow\ket{1}$. Readout is performed by reversing the initialization sequence, as described in the main text.}
\label{fig:main3}
\end{center}
\end{figure}

\paragraph*{Initialization and Readout}
Figures~\ref{fig:main3}(a)-(b) illustrate the initialization protocol. To prepare the system in state $\ket{\text{init}}=\ket{s_{LL}}\ket{s_{RR}}$ the detuning parameter $\varepsilon_{M}$ is set such that the (2,0,2) charge configuration is the ground state [Fig.~\ref{fig:main3}~(a)]. Due to the large single dot exchange splittings, the two electron ground state in the left and right dots is a singlet~\cite{Petta2005} and can be prepared with high fidelity~\cite{Atature2006}. Adiabatic tuning of $\varepsilon_{M}$ into a configuration where the (1,2,1) charge configuration is the ground state [Fig. 3(b)]  maps $\ket{\text{init}}\rightarrow\ket{1}$ through spin conserving tunneling events. Readout is performed in reverse, i.e., by detecting the tunneling of the electrons from the center dot to the left and right dots. Tunneling of $\ket{1}$ to the (2,0,2) ground state is allowed, but tunneling of $\ket{0}$ to the (2,0,2) ground state is Pauli blocked, similar to singlet-triplet readout in double dots~\cite{Petta2005}. Charge detection therefore provides a fast and high fidelity readout scheme. %We also note that the initialization and the readout are both performed (left-right) symmetrically since only the symmetric quadupolar detuning $\varepsilon_{M}$ needs to be tuned which allows to perform both operations at a charge noise sweet spot with respect to detuning noise.

\begin{figure}
\begin{center}
\includegraphics[width=1.\columnwidth]{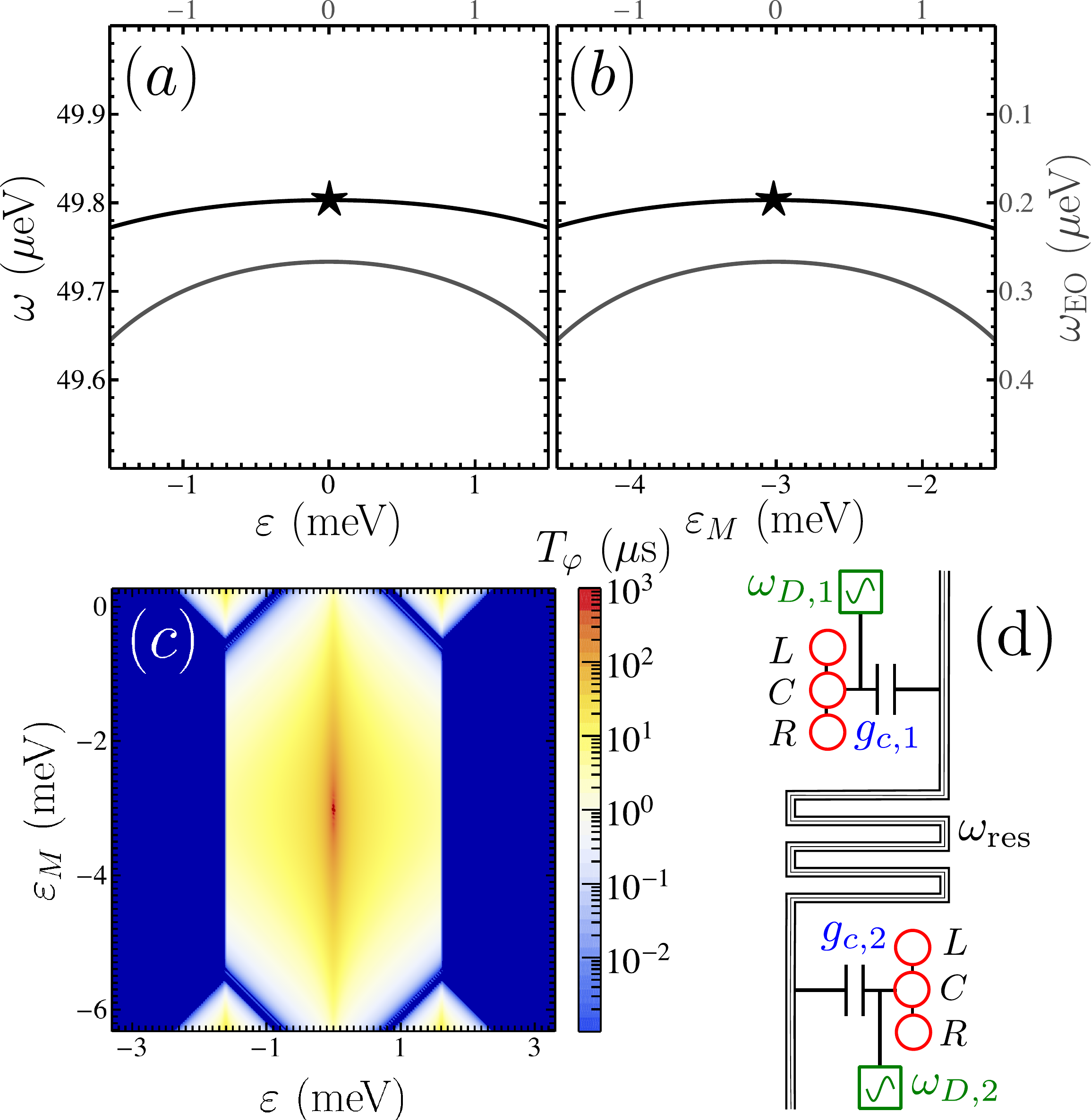}
\caption{ (a) Qubit splitting $\omega$ as a function of the (a) dipolar detuning $\varepsilon$ and (b) quadrupolar detuning $\varepsilon_{M}$. The stars in (a) and (b) mark the sweet spots where $\partial\omega_q/ \partial\varepsilon=0$ and $\partial\omega_q/ \partial\varepsilon_{M}=0$. For comparison the energy gap $\omega\st{EO}$ of the conventional exchange-only qubit (gray lines) is added with identical parameter settings that show steeper derivatives, thus, higher susceptibility to charge noise. (c) Qubit dephasing time $T_{\varphi}$ as a function of $\varepsilon$ and $\varepsilon_{M}$ for (c) $t^\prime_{l}=t^\prime_{r}$ using realistic parameters (see appendix). The simulation shows that $T_{\varphi}>\unit[100]{\mu s}$ if the qubit is operated at the charge noise sweet spot. (d) Schematic of the microwave resonator mediated two-qubit coupling. A cavity with resonance frequency $\omega\st{res}$ is coupled to the center dot with an effective charge-cavity coupling rate $g_{c,i}$. Each qubit is simultaneously driven at a frequency $\omega_{D,i}$.}
\label{fig:DSS}
\end{center}
\end{figure}

\paragraph*{Decoherence properties}
The Si/$\text{SiO}_{x}$ and Si/SiGe systems appear to be most favorable for the experimental realization of the QUEX qubit due to their large, and somewhat tunable, valley splitting. Since the four spin encoding of the QUEX is resilient to fluctuating LMFG noise~\cite{Sala2017} an implementation in natural silicon is still possible without significantly decreasing the decoherence time. The QUEX qubit possesses a full charge noise sweet spot where the qubit is first order insensitive to charge fluctuations in both detuning parameters $\varepsilon,\varepsilon_{M}$ [see Figs.~\ref{fig:DSS}~(a)-(b)]. From the condition $\partial_{\varepsilon}\omega_{q}=\partial_{\varepsilon_{M}}\omega_{q}=0$ and assuming $|t_{l}|=|t^\prime_{l}|$, $|t_{r}|=|t^\prime_{r}|$, and symmetric charging energies, we find a double sweet spot at $\varepsilon=0$ and $\varepsilon_{M}=(J_{c}-E_{V,C})/2-E_{C}$. Importantly, the sweet spot of the QUEX qubit is very flat compared to the conventional exchange-only qubit, i.e., reduced by  $1-k (E_{V}-J_{C})/E\st{charge}$ with the zero-bias splitting $E_{V}-J_{C}$ and $k>1$, therefore protecting it significantly better against charge noise. Numerical simulations shown in Fig.~\ref{fig:DSS}~(c) predictdephasing times on the scale of several hundreds of microseconds using realistic parameters, one order of magnitude larger than predictions for the conventional exchange-only qubit~\cite{Russ2016,Russ2017}. A full study for the general case $t_{l}\neq t^\prime_{l}$ and $t_{r}\neq t^\prime_{r}$ yields qualitatively similar results (see Sec.~\ref{app:gen}).

\paragraph*{Single-qubit operations}
Arbitrary single-qubit gates can be implemented by pulsing the exchange interactions, $J_{\mu}^{\nu}$. Assuming valley-orbit conserving and non-conserving tunneling to be equal ($t_{l}=t^\prime_{l}$ and $t_{r}=t^\prime_{r}$), the Hamiltonian~\eqref{eq:Hamqubit} can be rewritten as
\begin{align}
H_{q}=& \frac{1}{8}(2J_{0}^{l}-3J_{1}^{l}+2J_{0}^{r}-3J_{1}^{r}+8E_{V,C}-8J_{C})\sigma_{z} \nnb
&+\frac{1}{4}\frac{\sqrt{3}}{\sqrt{8}} (J_{0}^{l}+J_{1}^{l}-J_{0}^{r}-J_{1}^{r})\sigma_{x}.
\end{align}
Here, we introduced the Pauli operators $\sigma_{z}=\ket{0}\bra{0}-\ket{1}\bra{1}$ and $\sigma_{x}=\sigma_{+}+\sigma_{-}$.
Therefore, pulsing $J_{q}^{l}=J_{q}^{r}$ results in pure (and fast) $z$-rotations while pulses with $J_{q}^{l}\neq J_{q}^{r}$ yield rotations around a tilted axis~\cite{Hanson2007b}. Experimentally, the exchange interaction can either be controlled through detuning \cite{Petta2005} or tunnel barrier control~\cite{Bertrand2015,Reed2016,Martins2016,Malinowski2017}. Barrier control has the advantage that it operates the qubit at the charge noise sweet spot, thus, mitigating decoherence from charge noise during the exchange pulse. This statement, however, is only valid for $t_{l}\approx t^\prime_{l}$ and $t_{r}\approx t^\prime_{r}$ since otherwise the position of the sweet spot depends on the tunneling and moves during the qubit operation. In the latter case, one can combine barrier control with tilting control to compensate for the shift of the sweet spot. Similar pulsing schemes are commonly implemented to cancel crosstalk between electrostatic gates~\cite{Wiel2002}.

A more natural way of implementing single-qubit rotations in the QUEX qubit is to modulate the exchange coupling, $J_{\mu}^{\nu}\rightarrow J_{\mu}^{\nu}+j_{\mu}^{\nu}\cos\left(\omega_{D}+\phi\right)$ analogous to the resonant exchange (RX) qubit ~\cite{Medford2013,Taylor2013,Russ2017}. In a frame rotating with the drive frequency $\omega_{D}$, and neglecting fast oscillations at 2$\omega_{D}$, the Hamiltonian~\eqref{eq:Hamqubit} can be written in its eigenbasis as
\begin{align}
H_{q}=\frac{1}{2}(\omega_{q}-\omega_{D})\sigma_{z}+\frac{J_{x}^{1}}{2}\left(\E^{\I \phi}\ket{1}\bra{0}+\E^{-\I \phi}\ket{0}\bra{1}\right).
\end{align}
Here, the phase $\phi$ can be used to adjust the qubit rotation axis. Experimentally, this can realized by parametric modulation of either the detuning parameters, $\varepsilon,\varepsilon_{M}$, or the tunneling couplings, $t_{l,r}$ and $t^{\prime}_{l,r}$. The time for a Rabi flip is $\tau^{-1}_{x}\propto A_{q} \propto j_{x}\left[J_{x}\partial(J_{1}-J_{0})/\partial q-(J_{1}-J_{0})\partial J_{x}/\partial q\right]/(2\omega_{q})$, where $A_{q}$ denotes the amplitude of the driving of the parameter $\left. q\in\lbrace\varepsilon,\varepsilon_{M}, t_{l,r},t_{l,r}^{\prime}\rbrace\right.$. Using realistic parameters we find $A_{\varepsilon_{M}}\approx\unit[0.13]{\mu eV}$, thus, $\tau_{x}=\unit[30]{ns}$ while driving at the charge noise double sweet spot (see Sec.~\ref{sec:SQO}). Modulation of the tunnel barrier has the additional benefit of providing a dynamic sweet spot, where the Rabi drive is first-order insensitive to fluctuations in detuning, $\partial J^{1}_{x}/\partial q=0$. Recent experiments show that charge noise affecting the Rabi frequency significantly reduces the number of coherent exchange oscillations~\cite{Malinowski2017}. Thanks to the naturally large energy splitting of the qubit $\omega_{q}$ counter-rotating terms are small and off-resonant transitions are strongly suppressed, making strong driving feasible.

\paragraph*{Two-qubit interaction}
Exchange-only qubits are limited to short-ranged, exchange-based operations requiring complex pulse sequences~\cite{Nature2000,Bacon2000,Kempe2001,Bacon2003,Woodworth2006}. The QUEX qubit, with its large and tunable energy splitting, enables near-resonant coupling to high frequency resonators giving rise to new entanglement generation protocols~\cite{Reiter2012}. We now describe a resonantly-driven, cavity-mediated two-qubit entanglement protocol~\cite{Cirac1995,Childs2000,Wallraff2007,Leek2009,Srinivasa2016}. 

For this setup a superconducting cavity with resonance frequency $\omega\st{res}$ is capacitively coupled to the electrostatic potential $V_{C}$ at the center QD while $V_{C}$ of each qubit is simultaneously modulated with frequency $\omega_{D}\approx\omega_{q}$ and phase $\phi$ [see Fig.~\ref{fig:DSS}~(d)]. Thus, $V_{C}\rightarrow V_{C}(t)=V^{0}_{C}+V^{1}_{C}\cos(\omega_{D}+\phi)+g_{c}(a+a^{\dagger})$. Here $a^{\dagger}$ ($a$) creates (annihilates) a cavity photon with frequency $\omega\st{res}$, and $g_{c}\propto \sqrt{Z}$ is the charge-photon coupling strength and $Z$ is the characteristic impedance of the resonator~\cite{Samkharadze2016,Stockklauser2017,Landig2017,Samkharadze2017}. The effect of $V_{C}(t)$ on each qubit is described by the Hamiltonian (see Sec.~\ref{sec:twoQubit})
\begin{align}
H\st{int}=\omega\st{res}a^{\dagger}a+g\sigma_{x}(a+a^{\dagger}) + \Omega\cos(\omega_{D}+\phi)\sigma_{x},
\label{eq:qubitCavity}
\end{align}
where the second term describes the qubit-cavity interaction with coupling strength $g=g_{c}\bra{0}\partial_{V_{C}}H_{q}(V_{C})\ket{1}$ and the last term induces a spin-flip with Rabi frequency $\Omega=V^{1}_{C}\bra{0}\partial_{V_{C}}H_{q}(V_{C})\ket{1}$.
Following the protocol described in Ref.~\cite{Srinivasa2016}, Hamiltonian~\eqref{eq:qubitCavity} generates ``red'' and ``blue'' sideband transitions for the particular choice of $\Omega=\pm (\omega\st{res}-\omega_{D})$ and $\omega_{D}=\omega_{q}$
\begin{align}
H_{\mp}(g,\phi) =\frac{g}{2}\left(\E^{\pm\I\phi}a^{\dagger} \sigma_{\mp} +\E^{\mp\I\phi} a\sigma_{\pm}\right).
\end{align}
An entangling controlled-Z (CZ) gate is constructed using pulses of ``red'' and ``blue'' sideband transition gates $S_{\mp}(\phi,\tau)\equiv\exp[-\I t H_{\mp}(g,\phi)]$ combined with single-qubit rotations~\cite{Srinivasa2016}. Using experimentally feasible parameter settings, a CZ gate is possible within $\tau\approx\unit[340]{ns}$ (see appendix). For this implementation it is essential to be able to tune the qubit near resonance to fulfill $\Omega=\pm ( \omega\st{res}-\omega_{D})$ while simultaneously matching $\omega_{D}=\omega_{q}$, therefore, requiring a large and controllable qubit splitting.
Alternative two-qubit coupling schemes include exchange-based interaction~\cite{Bacon2003,Woodworth2006} and capacitative coupling~\cite{Koh2012,Pal2014}.

\paragraph*{Discussion}
In summary, we have proposed a quadrupolar exchange (QUEX) spin qubit that uses the spin of four electrons in a TQD and gives rise to a large controllable qubit splitting. Since the large energy gap suppresses the susceptibility to charge noise and the qubit can be fully operated at a charge noise sweet spot we predict dephasing times exceeding $\sim\unit[100]{\mu s}$ allowing for a high quality qubit implementation. A symmetric readout and initialization protocol can be used to perform fast and high fidelity measurements. Together with the proposed cavity-mediated, long-distance entangling protocol, these properties render the QUEX qubit suitable for implementation in a large-scale quantum information processing architecture.

\acknowledgments  
We acknowledge funding from ARO through Grant No. W911NF-15-1-0149 and the DFG through SFB 767. We thank M. Benito and J.M. Taylor  for helpful discussions.
\clearpage
\onecolumngrid
\appendix

\section{Derivation of the effective Hamiltonian for the quadrupolar exchange-only qubit}
\label{app:der}
In this section we present a derivation of the effective Hamiltonian which describes the setup discussed in the main text consisting of three linearly arranged quantum dots (QDs). The dots are occupied with a total of four electrons and are electrostatically tuned such that the (1,2,1) charge configuration is energetically favorable. Here, $(l,m,n)$ denotes the charge configuration with $l$ electrons in the left, $m$ electrons in the center, and $n$ electrons in the right dot. For our theoretical model we use an extended Hubbard model ($t-J$ model)~\cite[p.~25]{Auerbach1998} which includes intra-dot and inter-dot charging energies, the direct Coulomb exchange interaction between two electrons occupying the same dot with different valley-orbit quantum numbers, and a valley-orbit non-conserving tunneling between the dots (see section~\ref{ssec:beyHub}).

Our proposed system is described by the following Hamiltonian
\begin{align}
H=H_{0} + H\st{ch}+ H\st{tun} 
\label{eq:Hub}
\end{align}
where $H_{0}$ contains the quantum dot orbital energies $\epsilon_{i\alpha}$ and the chemical potentials $V_{i}$, $H\st{ch}$ the intra-dot (internal) charging energies $C_{i\alpha,i\beta}$, $K_{i\alpha,i\beta}$ and the inter-dot (external) charging energies $C_{i\alpha,j\beta}$, $K_{i\alpha,j\beta}$ with $i\neq j$, and $H\st{tun}$ the tunneling matrix elements $t_{i\alpha,j\beta}$,
%\begin{align}
%H_{0}&=\sum_{i,\alpha}(\epsilon_{i\alpha}+V_{i})(n_{i\alpha,\uparrow}+n_{i\alpha,\downarrow}),\label{eq:HamZero}\\
%H\st{ch,i}&=\sum_{i}\sum_{\alpha,\beta}\sum_{\sigma,\sigma^{\prime}}C^{\text{in}}_{i\alpha,i\beta}c^{\dagger}_{i\alpha,\sigma}c^{\dagger}_{i\beta,\sigma^{\prime}}c_{i\beta,\sigma^{\prime}}c_{i\alpha,\sigma}+K^{\text{in}}_{i\alpha,i\beta}\delta_{\sigma\sigma^{\prime}}c^{\dagger}_{i\alpha,\sigma}c^{\dagger}_{i\beta,\sigma^{\prime}}c_{i\alpha,\sigma}c_{i\beta,\sigma^{\prime}}\label{eq:HamOnsite}\\
%H\st{ch,e}&=\sum_{i\neq j}\sum_{\alpha\neq\beta}\sum_{\sigma,\sigma^{\prime}}C^{\text{ex}}_{i\alpha,j\beta}c^{\dagger}_{i\alpha,\sigma}c^{\dagger}_{j\beta,\sigma^{\prime}}c_{j\beta,\sigma^{\prime}}c_{i\alpha,\sigma}+K^{\text{ex}}_{i\alpha,j\beta}\delta_{\sigma\sigma^{\prime}}c^{\dagger}_{i\alpha,\sigma}c^{\dagger}_{j\beta,\sigma^{\prime}}c_{i\alpha,\sigma}c_{j\beta,\sigma^{\prime}},\label{eq:HamOffsite}\\
%H\st{tun}&=\sum_{i\neq j}\sum_{\alpha,\beta}\sum_{\sigma\sigma^{\prime}} t_{i\alpha,j\beta}\delta_{\sigma\sigma^{\prime}} c^{\dagger}_{i\alpha,\sigma} c_{i\beta,\sigma^{\prime}}.
%\label{eq:HamTun}
%\end{align}
\begin{align}
H_{0}&=\sum_{i,\alpha}(\epsilon_{i\alpha}+V_{i})(n_{i\alpha,\uparrow}+n_{i\alpha,\downarrow}),\label{eq:HamZero}\\
H\st{ch}&=\sum_{i,j}\sum_{\alpha,\beta}\sum_{\sigma,\sigma^{\prime}}\left(C_{i\alpha,j\beta}c^{\dagger}_{i\alpha,\sigma}c^{\dagger}_{j\beta,\sigma^{\prime}}c_{j\beta,\sigma^{\prime}}c_{i\alpha,\sigma}+K_{i\alpha,j\beta}\delta_{\sigma\sigma^{\prime}}c^{\dagger}_{i\alpha,\sigma}c^{\dagger}_{j\beta,\sigma^{\prime}}c_{i\alpha,\sigma}c_{j\beta,\sigma^{\prime}}\right),\label{eq:HamCharge}\\
H\st{tun}&=\sum_{i\neq j}\sum_{\alpha,\beta}\sum_{\sigma\sigma^{\prime}} t_{i\alpha,j\beta}\delta_{\sigma\sigma^{\prime}} c^{\dagger}_{i\alpha,\sigma} c_{i\beta,\sigma^{\prime}}.
\label{eq:HamTun}
\end{align}
Here, the indices $i,j=L,C,R$ refer to the QD, the indices $\alpha,\beta=v_{0},v_{1}$ to the valley-orbit, and $\sigma,\sigma^{\prime}=\uparrow,\downarrow$ label the electron spin. In this notation, $n_{i\alpha,\sigma}=c^{\dagger}_{i\alpha,\sigma} c_{i\alpha,\sigma}$ is the number operator and $c^{\dagger}_{i\alpha,\sigma}$ ($c_{i\alpha,\sigma}$) creates (annihilates) an electron in QD~$i$ (i=L,C,R) occupying orbital $\alpha$ ($\alpha=1,2$) with spin $\sigma$ ($\sigma=\uparrow,\downarrow$). 
$V_{i}$ is the dot potential affected by the electrostatic gates $V_{i}$, $E_{V,i}\equiv\epsilon_{iv_{0}}-\epsilon_{iv_{1}}$ is the valley-orbital splitting, $C_{i\alpha,i\beta}$ and $K_{i\alpha,i\beta}$ denote the onsite (internal) Coulomb and Coulomb-exchange energies of two electrons occupying the same dot, and $C_{i\alpha,j\beta}$ and $K_{i\alpha,j\beta}$ with $i\neq j$ describe the Coulomb and Coulomb-exchange interaction between electrons occupying different dots. These parameters, $C_{i\alpha,j\beta}$, $K_{i\alpha,j\beta}$, and $t_{i\alpha,j\beta}$ are given by computing the corresponding single- and two-particle matrix elements of the electronic wavefunctions (see subsection~\ref{ssec:beyHub}) or are experimentally accessed by fitting the parameters to transport or cavity measurements~\cite{Wiel2002,Burkard2016}. Note, that the above Hamiltonian directly enforces spin-conserving tunneling and Coulomb interaction through the Kronecker delta $\delta_{\sigma\sigma^{\prime}}$. This is justified by the relatively small spin-orbit interaction in silicon. Restricting ourselves to the situation where the (1,2,1) charge configuration is energetically favorable and considering large orbital spacings for the left and right dot, i.e. strong confinement of the electrons, we find that the excited valley-orbital levels in the outer two QDs are frozen out. Assuming symmetric confinement potentials and small dot sizes (charge distribution in different orbitals is small compared to the electron-electron repulsion), we find $U_{i}\equiv C_{i\alpha,i\beta}\approx C_{i\alpha^{\prime},i\beta^{\prime}}$, $U_{i,j}\equiv C_{i\alpha,j\beta}\approx C_{i\alpha^{\prime},j\beta^{\prime}}$ for $\alpha,\alpha^{\prime},\beta,\beta^{\prime}=v_{0},v_{1}$, and $J_{C}\equiv K_{iv_{0},iv_{1}}=K^{\text{in}}_{iv_{1},iV_{0}}$ while $K_{i\alpha,i\alpha}=0$. To introduce the same notation as used in the main paper, we now omit the orbital index if $\alpha=v_{0}$ and use $Q^{\prime}$ if $\alpha=v_{1}$ for $Q=L,C,R$. Hence, there are four relevant tunneling amplitudes in the regime of our interest
\begin{align}
t_{l}\equiv& t_{L,C} + K_{L,C} = t_{C,L}^{*} + K_{C,L}^{*},\\
t^{\prime}_{l}\equiv& t_{L,C^{\prime}} + K_{L,C^{\prime}} = t_{C^{\prime},L}^{*} + K_{C^{\prime},L}^{*},\\
t_{r}\equiv& t_{C,R} + K_{C,R} = t_{R,C}^{*} + K_{R,C}^{*},\\
t^{\prime}_{r}\equiv& t_{C^{\prime},R} + K_{C^{\prime},R} = t_{R,C^{\prime}}^{*} + K_{R,C^{\prime}}^{*},
\end{align}
where $t_{l,(r)}$ and $t^{\prime}_{l,(r)}$ describe valley-orbit conserving and non-conserving tunneling between the left (right) QD and the center QD renormalized by the Coulomb-exchange interaction. In a linear array the tunneling matrix element $t_{13}\equiv t_{L,R} + K_{L,R} = t_{R,L}^{*} + K_{R,L}^{*}\approx 0$ between the left and right dot is small and neglected for all analytical calculations in this supplement. 

For later convenience, we also define the dipolar-detuning, $\varepsilon=(V_{L}-V_{R})/2$, and the quadrupolar detuning, $\varepsilon_{M}=V_{C}-(V_{L}+V_{R})/2$. An applied magnetic field sets the quantization axis ($z$) of the qubit and allows for a separate analysis of each subspace spanned by the states with total spin $S=|\boldsymbol{S}|\equiv|\sum_{\mu}\boldsymbol{S}_{\mu}|$ and $S_{z}$. Note, that the $z$-projection of the spin is given by the direction of the magnetic field and not by the geometric coordinate system $x,y,z$ of the quantum dot system.

The quadrupolar exchange-only spin qubit is implemented in the (1,2,1) charge configuration using the logical spin qubit states~\cite{Lidar2012,Sala2017}
\begin{align}
\ket{0}=&\ket{s_{LR}}\ket{s_{CC}}\nonumber\\
	  =&\frac{1}{\sqrt{2}}\left(c^{\dagger}_{Lv_{0},\uparrow}c^{\dagger}_{Cv_{0},\uparrow}c^{\dagger}_{Cv_{0},\downarrow}c^{\dagger}_{Rv_{0},\downarrow}-c^{\dagger}_{Lv_{0},\downarrow}c^{\dagger}_{Cv_{0},\uparrow}c^{\dagger}_{Cv_{0},\downarrow}c^{\dagger}_{Rv_{0},\uparrow}\right)\ket{\text{vac}}
\label{eq:state0app}\\
\ket{1}=&\frac{1}{\sqrt{3}}(\ket{s_{LC^{\prime}}}\ket{s_{CR}}+\ket{s_{LC}}\ket{s_{C^{\prime}R}})\nonumber\\
	  =&\frac{1}{2\sqrt{3}}\Big(
	  c^{\dagger}_{Lv_{0},\uparrow}c^{\dagger}_{Cv_{0},\uparrow}c^{\dagger}_{Cv_{1},\downarrow}c^{\dagger}_{Rv_{0},\downarrow}+
	  c^{\dagger}_{Lv_{0},\uparrow}c^{\dagger}_{Cv_{0},\downarrow}c^{\dagger}_{Cv_{1},\uparrow}c^{\dagger}_{Rv_{0},\downarrow}\nonumber\\ &\phantom{\frac{1}{2\sqrt{3}}}
	  -2c^{\dagger}_{Lv_{0},\uparrow}c^{\dagger}_{Cv_{0},\downarrow}c^{\dagger}_{Cv_{1},\downarrow}c^{\dagger}_{Rv_{0},\uparrow}-
	  2c^{\dagger}_{Lv_{0},\downarrow}c^{\dagger}_{Cv_{0},\uparrow}c^{\dagger}_{Cv_{1},\uparrow}c^{\dagger}_{Rv_{0},\downarrow}\nonumber\\ &\phantom{\frac{1}{2\sqrt{3}}}
	  +c^{\dagger}_{Lv_{0},\downarrow}c^{\dagger}_{Cv_{0},\uparrow}c^{\dagger}_{Cv_{1},\downarrow}c^{\dagger}_{Rv_{0},\uparrow}+
	  c^{\dagger}_{Lv_{0},\downarrow}c^{\dagger}_{Cv_{0},\downarrow}c^{\dagger}_{Cv_{1},\uparrow}c^{\dagger}_{Rv_{0},\uparrow}
	  \Big)\ket{\text{vac}}
\label{eq:state1app}
\end{align}
both residing in the $S=0$ subspace of the four electron system and forming a decoherence-free subspace (DFS) qubit~\cite{Nature2000,Lidar2012}. Here, $\ket{\text{vac}}$ denotes the vacuum state, $\ket{s_{\mu\nu}}=(\ket{\uparrow}_{\mu}\ket{\downarrow}_{\nu}-\ket{\downarrow}_{\mu}\ket{\uparrow}_{\nu})\sqrt{2}$ denotes the spin singlet state between electrons in orbitals $\mu$ and $\nu$ with $\mu,\nu=L,C,C^{\prime},R$ where $L$ ($R$) reside in the left (right) dot and $C,(C^{\prime})$ reside in the lower (upper) orbital in the center dot. There are two additional (leakage) states with the same total spin and charge configuration
\begin{align}
\ket{0^{+}}&=\ket{s_{LR}}\ket{s_{CC^{\prime}}},\label{eq:statesP}\\
\ket{0^{++}}&=\ket{s_{LR}}\ket{s_{C^{\prime}C^{\prime}}}
\label{eq:statesPP}
\end{align}
occupying excited valley-orbit levels in the center dot. There are no direct hopping matrix elements between the states $ \ket{0}$, $\ket{1}$, $\ket{0^{+}}$, and $\ket{0^{++}}$ and their Hamiltonian is given in the basis $\lbrace \ket{0},\ket{1},\ket{0^{+}},\ket{0^{++}}\rbrace$ by the diagonal matrix
\begin{align}
H^{0}_{(1,2,1)}=\text{diag}(0,E_{V,C}-J_{c},E_{V,C},2E_{V,C}).
\end{align}

However, coherent tunneling of the electrons couples states with (1,2,1) charge configuration to states with different charge occupations, i.e, states with (2,2,0), (2,1,1), (1,3,0), (0,3,1), (1,1,2) and (0,2,2) charge configuration, each energetically separated at least by the non-local nearest neighbor Coulomb repulsion $U_{ij}$. Considering only the lowest energy states within these charge configurations, we have to take the following 10 states with their respective energy into consideration;
%\begin{align}
%\ket{(2,1,1)_{1}}&=\ket{\uparrow_{A}}\ket{\downarrow_{A}}\ket{s_{CR}}, \hfill E_{4}=\varepsilon-\varepsilon_{M}+U_{1}-U_{2}+U_{2,3},\\
%\ket{(2,1,1)_{2}}&=\ket{\uparrow_{A}}\ket{\downarrow_{A}}\ket{s_{C^\prime R}}, \hfill E_{5}=\varepsilon-\varepsilon_{M}+U_{1}-U_{2}+U_{2,3}+E_{V,C},\\
%\ket{(1,1,2)_{1}}&=\ket{s_{LC}}\ket{\uparrow_{D}}\ket{\downarrow_{D}},\hfill E_{6}=-\varepsilon-\varepsilon_{M}+U_{3}-U_{2}+U_{1,2},\\
%\ket{(1,1,2)_{2}}&=\ket{s_{LC^\prime}}\ket{\uparrow_{D}}\ket{\downarrow_{D}},\hfill E_{7}=-\varepsilon-\varepsilon_{M}+U_{3}-U_{2}+U_{1,2}+E_{V,C},\\
%\ket{(1,3,0)_{1}}&=\ket{s_{LC^\prime}}\ket{\uparrow_{B}}\ket{\downarrow_{B}},\hfill E_{8}=\varepsilon+\varepsilon_{M}+2U_{2}-U_{1,2}+U_{2,3}+E_{V,C}-J_{C},\\
%\ket{(1,3,0)_{2}}&=\ket{s_{LC}}\ket{\uparrow_{C}}\ket{\downarrow_{C}},\hfill E_{9}=\varepsilon+\varepsilon_{M}+2U_{2}-U_{1,2}+2U_{2,3}+2E_{V,C}-J_{C}\\
%\ket{(0,3,1)_{1}}&=\ket{\uparrow_{B}}\ket{\downarrow_{B}}\ket{s_{C^\prime R}},\hfill E_{10}=\varepsilon_{M}-\varepsilon+2U_{2}+2U_{1,2}-U_{2,3}+E_{V,C}-J_{C}\\
%\ket{(0,3,1)_{2}}&=\ket{s_{CR}}\ket{\uparrow_{C}}\ket{\downarrow_{C}},\hfill E_{11}=\varepsilon_{M}-\varepsilon+2U_{2}+2U_{1,2}-U_{2,3}+2E_{V,C}-J_{C}\\
%\ket{(2,2,0)}&=\ket{\uparrow_{A}}\ket{\downarrow_{A}}\ket{s_{C C^\prime}},\,\,\, E_{12}=2\varepsilon+U_{1}-2U_{1,2}+U_{2,3}+E_{V,C}+J_{C}\\
%\ket{(2,2,0)}&=\ket{s_{C C^\prime}}\ket{\uparrow_{D}}\ket{\downarrow_{D}},\,\, E_{13}=-2\varepsilon+U_{3}+2U_{1,2}-U_{2,3}+E_{V,C}+J_{C},
%\end{align}
\begin{align}
\ket{(2,1,1)_{1}}&=\ket{s_{LL}}\ket{s_{CR}},\, \text{with}\, E_{4}=\varepsilon-\varepsilon_{M}+U_{1}-U_{2}+U_{2,3},\label{eq:states0}\\
\ket{(2,1,1)_{2}}&=\ket{s_{LL}}\ket{s_{C^\prime R}}, \, \text{with}\, E_{5}=\varepsilon-\varepsilon_{M}+U_{1}-U_{2}+U_{2,3}+E_{V,C},\\
\ket{(1,1,2)_{1}}&=\ket{s_{LC}}\ket{s_{RR}},\, \text{with}\, E_{6}=-\varepsilon-\varepsilon_{M}+U_{3}-U_{2}+U_{1,2},\\
\ket{(1,1,2)_{2}}&=\ket{s_{LC^\prime}}\ket{s_{RR}},\, \text{with}\, E_{7}=-\varepsilon-\varepsilon_{M}+U_{3}-U_{2}+U_{1,2}+E_{V,C},\\
\ket{(1,3,0)_{1}}&=\ket{s_{LC^\prime}}\ket{s_{CC}},\, \text{with}\, E_{8}=\varepsilon+\varepsilon_{M}+2U_{2}-U_{1,2}+U_{2,3}+E_{V,C}-J_{C},\\
\ket{(1,3,0)_{2}}&=\ket{s_{LC}}\ket{s_{C^\prime C^\prime}},\, \text{with}\, E_{9}=\varepsilon+\varepsilon_{M}+2U_{2}-U_{1,2}+2U_{2,3}+2E_{V,C}-J_{C}\\
\ket{(0,3,1)_{1}}&=\ket{s_{CC}}\ket{s_{C^\prime R}},\, \text{with}\, E_{10}=\varepsilon_{M}-\varepsilon+2U_{2}+2U_{1,2}-U_{2,3}+E_{V,C}-J_{C}\\
\ket{(0,3,1)_{2}}&=\ket{s_{CR}}\ket{s_{C^\prime C^\prime}},\, \text{with}\, E_{11}=\varepsilon_{M}-\varepsilon+2U_{2}+2U_{1,2}-U_{2,3}+2E_{V,C}-J_{C}\\
\ket{(2,2,0)}&=\ket{s_{LL}}\ket{s_{C C^\prime}},\, \text{with}\,E_{12}=2\varepsilon+U_{1}-2U_{1,2}+U_{2,3}+E_{V,C}+J_{C}\\
\ket{(0,2,2)}&=\ket{s_{C C^\prime}}\ket{s_{RR}},\, \text{with}\,E_{13}=-2\varepsilon+U_{3}+2U_{1,2}-U_{2,3}+E_{V,C}+J_{C},
\label{eq:states}
\end{align}
In the remainder of the derivation of the effective Hamiltonian we neglect the influence of $\ket{(2,2,0)}$ and $\ket{(0,2,2)}$ since they only contribute terms of the order $\sim t_{l,(L)}^{2}t_{r,(R)}^{2}$~\cite{Russ2017}. However, we take these terms into consideration for numerical calculations.

The effective low-energy subspace in the (1,2,1) charge regime can be approximated via a Schrieffer-Wolff transformation in the limit $t_{l,r},t_{L,R}\ll | U_{i}-U_{i,j}\pm\varepsilon\pm\varepsilon_{M}|$. The resulting effective Hamiltonian in the (1,2,1) subspace with $S=S_{z}=0$ is given as follows
\begin{align}
H\st{eff}=\left(
\begin{array}{cccc}
 J_{0,0} & \tilde{J}_{0,1}& \tilde{J}_{0,2} & 0 \\
\tilde{J}_{0,1}^* & J_{1,1}+E_{V,C}-J_{C} &J_{1,2} & \tilde{J}_{1,3} \\
\tilde{J}_{0,2}^* & J_{1,2} & J_{2,2}+E_{V,C} & \tilde{J}_{2,3} \\
 0 & \tilde{J}_{1,3}^* & \tilde{J}_{2,3}^* & J_{3,3}+2E_{V,C} \\
\end{array}
\right).
\label{eq:HamSub}
\end{align}
For the relevant exchange couplings we find the following analytic expressions
\begin{align}
J_{0,0}=&-\frac{1}{4}\left(\frac{|t_{l}|^{2}}{E_{4}}+\frac{|t_{r}|^{2}}{E_{6}}+\frac{|t_l^{\prime}|^{2}}{E_{10}}+\frac{|t_r^{\prime}|^{2}}{E_{8}}\right),\\
\tilde{J}_{0,1}=&-\frac{1}{4} \sqrt{\frac{3}{8}} \left(\frac{   t_{l}^*  t_l^{\prime}}{E_{4}}+\frac{   t_{l}^*  t_l^{\prime}}{E_{4}+J_{C}-E_{V,C}} +\frac{   t_{l}^*  t_l^{\prime}}{E_{10}}+\frac{   t_{l}^*  t_l^{\prime}}{E_{10}+J_{C}-E_{V,C}}\right.\nnb
&\phantom{-\frac{1}{4} \sqrt{\frac{3}{8}}}-\left. \frac{t_r^{\prime*}  t_{r}}{E_{6}}-\frac{   t_r^{\prime*}  t_{r}}{E_{6}+J_{C}-E_{V,C}}-\frac{   t_r^{\prime*}  t_{r}}{E_{8}}-\frac{   t_r^{\prime*}  t_{r}}{E_{8}+J_{C}-E_{V,C}}\right),\\
J_{1,1}=&-\frac{3}{8} \left(\frac{|t_l|^2}{E_{5}+J_{C}-E_{V,C}}+\frac{|t_l|^2}{E_{10}+J_{C}-E_{V,C}}+\frac{|t_l^{\prime}|^2}{E_{4}+J_{C}-E_{V,C}}+\frac{|t_l^{\prime}|^2}{E_{11}+J_{C}-E_{V,C}}\right.\nnb
&\left.\quad+\frac{|t_r|^2}{E_{7}+J_{C}-E_{V,C}}+\frac{|t_r|^2}{E_{8}+J_{C}-E_{V,C}}+\frac{|t_r^{\prime}|^2}{E_{6}+J_{C}-E_{V,C}}+\frac{|t_r^{\prime}|^2}{E_{9}+J_{C}-E_{V,C}}\right),
\end{align}
while for the higher excitation couplings we refer to Section~\ref{sec:highEx}. We note, that $J_{i,j}=J_{i,j}^{l}+J_{i,j}^{l^{\prime}}+J_{i,j}^{r}+J_{i,j}^{r^{\prime}}$, where $J_{i,j}^{q}\propto |t_{q}|^{2}$ are real parameters and describe a Heisenberg-type exchange interaction between states occupying the same valley, while $\tilde{J}_{i,j}=\tilde{J}_{l,(i,j)}+\tilde{J}_{r,(i,j)}$, where $\tilde{J}_{p,(i,j)}\propto t_{p}^{*}t_{p^{\prime}}$ with $p=l,r$ and $p^{\prime}=l^{\prime},r^{\prime}$ are complex parameters and describe valley-orbit non-conserving exchange.
Since the degeneracy of the (1,2,1) charge states is lifted due to the valley-orbit splitting $E_{V,C}$ and the intra-dot direct Coulomb exchange $J_{c}$, we use a second Schrieffer-Wolff transformation assuming $J_{i,j}\ll ||E_{V,C}-J_{c}|-E_{V,C}|$ to find the dynamics in the qubit subspace $\lbrace \ket{1},\ket{0}\rbrace$. Defining
\begin{align}
J_{0}\approx& J_{0,0}+\frac{|\tilde{J}_{0,2}|^{2}}{J_{0,0}-J_{2,2}}\\
J_{1}\approx& J_{1,1}+\frac{|J_{1,2}|^{2}}{J_{1,1}-J_{2,2}}+\frac{|\tilde{J}_{1,3}|^{2}}{J_{1,1}-J_{3,3}}\\
J_{x}\approx& \tilde{J}_{0,1}+\frac{\tilde{J}_{0,2}J_{1,2}(J_{0,0}+J_{1,1}-2J_{2,2})}{2(J_{0,0}-J_{2,2})(J_{1,1}-J_{2,2})}
\end{align}
we find for the qubit Hamiltonian the same expressions as used in the main text,
\begin{align}
H\st{q}= J_{0}\ket{0}\bra{0}+(J_{1}+E_{V,C}-J_{C})\ket{1}\bra{1} + J_{x} \ket{1}\bra{0}+ J_{x}^{*} \ket{0}\bra{1}
\label{eq:Hamqubit}
\end{align}
with the qubit splitting
\begin{align}
\omega_{q}=\sqrt{(J_{1}+E_{V,C}-J_{C}-J_{0})^{2}+|J_{x}|^{2}}\approx (J_{1}+E_{V,C}-J_{C}-J_{0})+\frac{|J_{x}|^{2}}{2(J_{1}+E_{V,C}-J_{C}-J_{0})}.
\end{align}
We note that $J_{x}$ couples the two logical qubit states $\ket{0}$ and $\ket{1}$ and is tunable through electrical control of the tunnel barrier or the detuning parameters. Moreover, $J_{1}\sim E_{V,C}-J_{C}\gg J_{x}$ is dominated by the valley-orbit splitting and Coulomb exchange, therefore ``always-on'', giving rise to a large zero-bias splitting.

\section{Molecular orbital analysis}
\label{ssec:beyHub}

\begin{figure}
\begin{center}
\includegraphics[width=0.6\columnwidth]{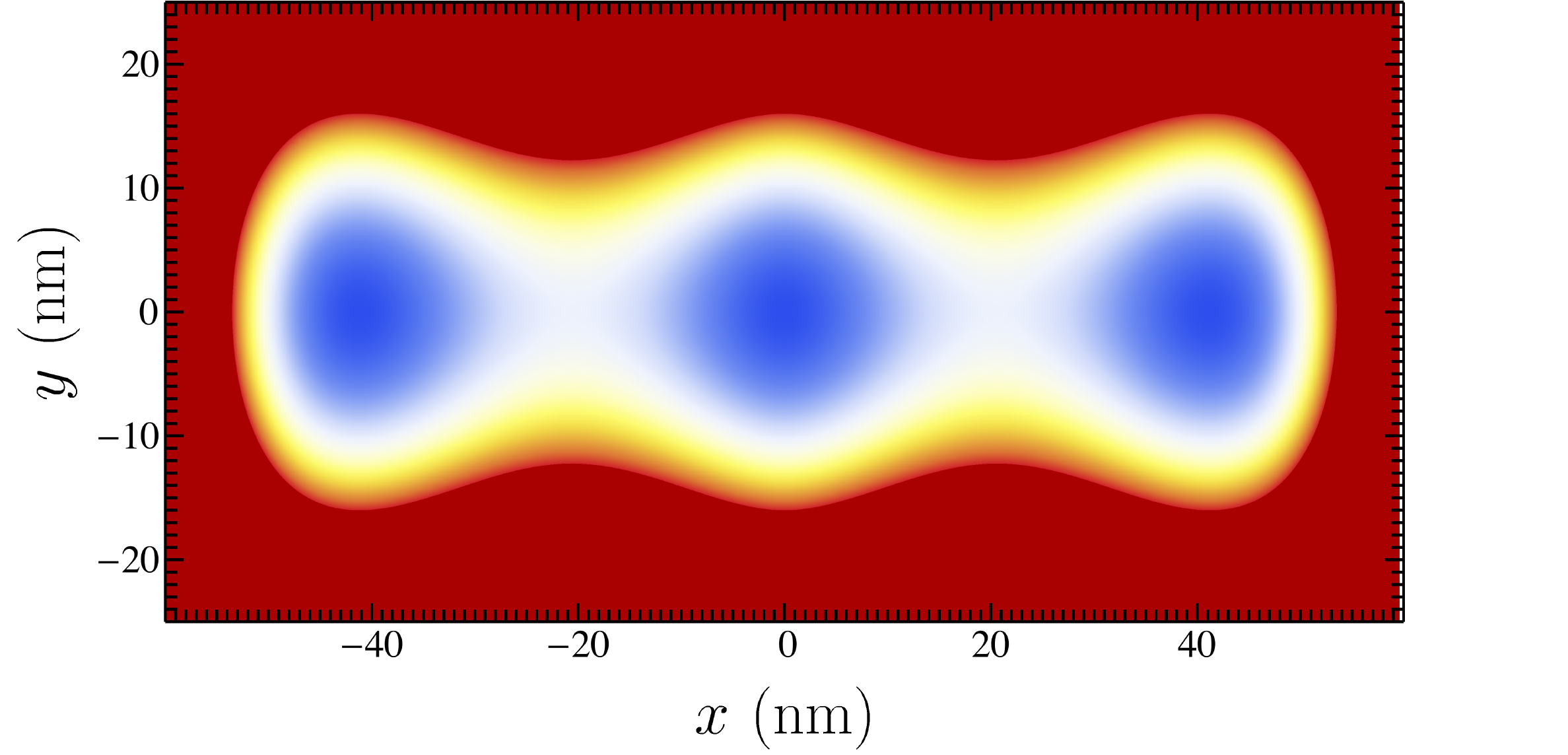}
\caption{Modeling of the triple quantum dot in the 2DEG using the harmonic square wells of Eq.~\eqref{eq:pot}. The $x$-axis is aligned along the triple dot axis, the $y$-axis is perpendicular in the 2DEG plane, and the $z$-axis is perpendicular to the plane. The following parameters have been chosen: $a_{l}=a_{r}=\unit[41.2]{nm}$, $c_{l}=c_{r}=0$, and $E\st{orb}=\unit[6]{meV}$.}
\label{fig:pot}
\end{center}
\end{figure}

In this section we support our analysis using a full Hund-Mullikan orbital calculation. Following Ref.~\cite{Burkard1999,Russ2015b}, the triple quantum dot is modeled by a potential of the form 

\begin{align}
\mathcal{V}(x,y) = \left\lbrace\begin{array}{cc}
 \frac{m\omega\st{dot}^{2}}{2}\left[-\frac{1}{48} a_l^2 c_l \left(c_l^3-6 c_l-8\right)+\frac{x^4}{a_l^2}-\frac{2 (c_l-3) x^3}{3 a_l}-(c_l-1) x^2\right] & x<0 \\
  \frac{m\omega\st{dot}^{2}}{2}\left[-\frac{1}{48} a_l^2 c_l \left(c_l^3-6 c_l-8\right)+\frac{x^4}{a_r^2}-\frac{2 (c_r+3) x^3}{3 a_r}+(c_r+1) x^2\right] & x\ge 0 \\
\end{array}\right. 
\label{eq:pot}
\end{align} 
which separates into three harmonic wells of frequency $\hbar\omega\st{dot}=E\st{orb}$ (see Fig.~\ref{fig:pot}).
Here, $a_{l,(r)}$ is the distance between the left (right) and the center quantum dot and the parameter $|c_{l,(r)}|<1$ phenomenologically describes the slope in energy between the left (right) dot and center dot. The $c_{l,r}$ are related to the detuning parameters by
%\begin{align}
%\varepsilon=-\frac{\omega\st{dot}}{6}(a_{r}^{2}c_{r}+a_{l}^{2}c_{l}),\\
%\varepsilon_{M}=-\frac{\omega\st{dot}}{6}(a_{r}^{2}c_{r}-a_{l}^{2}c_{l}),
%\end{align}
\begin{align}
c_{l}=-\frac{6(\varepsilon-\varepsilon_{M})}{a_{l}^{2}m\omega\st{dot}^{2}},\label{eq:cl}\\
c_{r}=-\frac{6(\varepsilon+\varepsilon_{M})}{a_{r}^{2}m\omega\st{dot}^{2}}.\label{eq:cr}
\end{align}
In this model, the height of the tunnel barrier $h_{l,r}$ is given by the inter-dot distance $h_{l,r}\propto a_{l,r}$.

In the effective mass (EM) approximation the orbital ground state electron wavefunctions of a multi-valley system with valley $\xi=z,\bar{z}$ residing in quantum dot $q=l,c,r$ is given by~\cite{Saraiva2009,Culcer2009,Hu2011,Saraiva2011,Gamble2012,Wu2012,Zimmerman2017}
\begin{align}
\chi^{\xi s}_q(x,y,z)&=\Phi^{s}\st{orb}(x,y)\times\Psi(z)\times u_{\xi}(x,y,z)\times\E^{i k_{\xi}z}\nonumber\\
&\approx\frac{\E^{\frac{i a_{q} y}{2 l_{B}^2}} \E^{-\frac{(x-a_{q})^2+y^2}{2 R_{q}^2}}}{ \sqrt{\pi} R_{q}}\times \frac{\E^{-\frac{z^2}{2 d^2}} }{\sqrt{2}\pi ^{1/4} \sqrt{d}}\times 1 \times\E^{i k_{\xi}z}.
\label{eq:wave0s}
\end{align}
Note, that setting $u_{\xi}(x,y,z)=\text{const}$ does not affect our results since all relevant energies vary slowly with respect to the lattice site, therefore, all matrix elements are averaged over several periods of the Bloch function.
Here, the first factor describes the orbital  confinement in the $xy$-plane at $x$-positions $a_{l},a_{c}=0,a_{r}$ with quantum dot radius $R_{q}=\hbar/\sqrt{m E_{\text{orb},q}}<a_{l},a_{r}$, and the magnetic length $l_{B}=\sqrt{\hbar/eB}$. The remaining factors describe the confinement along $z$-direction in a quantum well with length $d\ll R_{q}$, with $z$-valley position $k_{\xi}=\pm k_{0}$ in $k$-space. The electronic quantum dot wavefunction of a multi-valley system in the first excited orbital state can similarly be approximated by~\cite{Friesen2007,Saraiva2009,Culcer2009,Friesen2010,Culcer2010,Culcer2010b,Wu2012,Tahan2014,Zimmerman2017}.
\begin{align}
\chi^{\xi p}_q(x,y,z)&=\Phi^{p}\st{orb}(x,y)\times\Psi(z)\times u_{\xi}(x,y,z)\times\E^{i k_{\xi}z}\nonumber\\
&\approx\left[(x-a_{q})+\I y\right]\frac{\E^{\frac{i a_{q} y}{2 l_{B}^2}} \E^{-\frac{(x-a_{q})^2+y^2}{2 R_{q}^2}}}{ \sqrt{\pi} R_{q}^{2}}\times \frac{\E^{-\frac{z^2}{2 d^2}} }{\sqrt{2}\pi ^{1/4} \sqrt{d}}\times 1 \times\E^{i k_{\xi}z}.
\label{eq:wave1s}
\end{align}
In realistic devices, however, miscuts, atomistic steps at the interface, or other effects couple the valley and orbital degrees of freedom. In general, the valley-orbit coupling can be described  in the basis of each dot $\lbrace \chi^{zs}_{q},\chi^{\bar{z}s}_{q},\chi^{zp}_{q},\chi^{\bar{z}p}_{q}\rbrace$ by~\cite{Friesen2010,Rancic2016,Zimmerman2017}
\begin{align}
H\st{VO,q}=\frac{1}{2}\left(
\begin{array}{cccc}
 0 & \Delta_{\text{VO}s,q}& \Delta_{Oz,q} & \Delta_{Vs,q} \\
\Delta_{\text{VO}s,q}^* & 0 &\Delta_{Vp,q} & \Delta_{O\bar{z},q} \\
\Delta_{Oz,q}^* & \Delta_{Vp,q}^{*} & 2E_{\text{orb},q} & \Delta_{\text{VO}p,q} \\
\Delta_{Vs,q}^{*} &  \Delta_{O\bar{z},q}^* & \Delta_{\text{VO}p,q}^* & 2E_{\text{orb},q} \\
\end{array}
\right)=H\st{VO,q}^{0}+H\st{VO,q}^{1},
\label{eq:VOmatrix}
\end{align}
where $H\st{VO,q}^{0}$ contains only the diagonal elements and $H\st{VO,q}^{1}$ only off-diagonal elements.
Here, $\Delta_{\text{VO}s}$ ($\Delta_{\text{VO}p}$) is the intra-orbit inter-valley coupling of the ground (excited) state, $\Delta_{O,\xi} $ denotes the inter-orbit intra-valley coupling of the $\xi=z,\bar{z}$ valley, and $\Delta_{Vs}$ ($\Delta_{Vp}$) describes the inter-orbit inter-valley coupling between the orbital ground (excited) states.
The eigenstates of the Hamiltonian~\eqref{eq:VOmatrix} are calculated in two steps. First, a SW transformation is applied to block-diagonalize $\widetilde{H}\st{VO,q}=\E^{S}H\st{VO,q}\E^{-S}$ with $|\Delta_{Vs},\Delta_{Vp},\Delta_{Oz},\Delta_{O\bar{z}}|\ll E_{\text{orb},q}$. Thus, we obtain (up to a energy shift)
 \begin{align}
\widetilde{H}\st{VO,q}\approx\frac{1}{2}\left(
\begin{array}{cccc}
 -2E_{s,q} & \widetilde{\Delta}_{\text{VO}s,q}& 0 & 0 \\
\widetilde{\Delta}_{\text{VO}s,q}^* & 2E_{s,q} &0 & 0 \\
0 & 0 & 2(E_{\text{orb},q}-E_{p,q}) & \widetilde{\Delta}_{\text{VO}p,q} \\
0 &  0 & \widetilde{\Delta}_{\text{VO}p,q}^* & 2(E_{\text{orb},q}+E_{p,q}) \\
\end{array}
\right),
\label{eq:VOmatrixSW}
\end{align}
in the basis of the valley-orbit mixed states $\ket{\varphi^{\xi o}_{q}}= \E^{-S}\ket{\chi^{\xi o}_{q}}$ with $o=s,p$ and $\xi = z,\bar{z}$. The anti-hermitian matrix $S$ is set by the SW condition $[S,H\st{VO,q}^{0}]=-H\st{VO,q}^{1}$. Note, that in general the mixing between $s$- and $p$-orbitals is different for $z$ and $\bar{z}$ valley which we will find useful below. The complex valley-orbit coupling of the ground state $\widetilde{\Delta}_{\text{VO}s,q}=|\widetilde{\Delta}_{\text{VO}s,q}|\E^{i \phi^{v}_{q}}$ now defines the valley-splitting $h_{V,q}=|\widetilde{\Delta}_{\text{VO}s,q}|$ and the phase of the valley pseudo-spin $\phi^{v}_{q}$ (similarly for the $p$-orbitals). The orbital splittings $E_{s,q},E_{p,q}\ll |\widetilde{\Delta}_{\text{VO}s,q}|,|\widetilde{\Delta}_{\text{VO}p,q}|$ are neglected in the remainder of the supplement.
 
In a second step, we diagonalize the $s$-orbital $2\times 2$-block using the basis transformation $\ket{\varphi^{\xi s}_{q}}\rightarrow (\ket{\varphi^{\xi s}_{q}}\mp\E^{i \phi^{v}_{q}}\ket{\varphi^{\bar{\xi} s}_{q}})/\sqrt{2}$ and similarly for the $p$-orbitals.
The electron wavefunctions are then given by

\begin{align}
\ket{\varphi^{0s}_q}=&(\E^{-S}\ket{\chi^{z s}_{q}}-\E^{i \phi^{v}_{q}}\E^{-S}\ket{\chi^{\bar{z} s}_{q}})/\sqrt{2},
\\
\ket{\varphi^{1s}_q}=&(\E^{-S}\ket{\chi^{z s}_{q}}+\E^{i \phi^{v}_{q}}\E^{-S}\ket{\chi^{\bar{z} s}_{q}})/\sqrt{2},
\\
\ket{\varphi^{0p}_q}=&(\E^{-S}\ket{\chi^{z p}_{q}}-\E^{i \phi^{v}_{p,q}}\E^{-S}\ket{\chi^{\bar{z} p}_{q}})/\sqrt{2},
\\
\ket{\varphi^{1p}_q}=&(\E^{-S}\ket{\chi^{z p}_{q}}+\E^{i \phi^{v}_{p,q}}\E^{-S}\ket{\chi^{\bar{z} p}_{q}})/\sqrt{2},
\\
\end{align}
with $\phi^{v}_{p,q}=\arg(\widetilde{\Delta}_{\text{VO}p,q})$ being the valley pseudo-spin phase of the $p$-orbital.
The set of wavefunctions introduced above $\boldsymbol{\varphi}\equiv\lbrace\varphi^{0s}_{l},\varphi^{0s}_{c},\varphi^{0s}_{r},\varphi^{1s}_{l},\varphi^{1s}_{c},\varphi^{1s}_{r},\varphi^{0p}_{l},\varphi^{0p}_{c},\varphi^{0p}_{r},\varphi^{1p}_{l},\varphi^{1p}_{c},\varphi^{1p}_{r}\rbrace^{T}$, however, is unsuitable for further calculations due to finite overlaps
\begin{align}
\Sigma_{ij}^{\alpha\beta}=  \int d^{3}\boldsymbol{r} \varphi^{\alpha*}_{i}(\boldsymbol{r}) \varphi^{\beta}_{j}(\boldsymbol{r})\neq \delta_{ij}\delta_{\alpha\beta}.
\end{align}
Enforcing the orthonormalization conditions yields a new set of wavefunctions \\$\boldsymbol{\Phi}\equiv\lbrace\Phi^{0s}_{l},\Phi^{0s}_{c},\Phi^{0s}_{r},\Phi^{1s}_{l},\Phi^{1s}_{c},\Phi^{1s}_{r},\Phi^{0p}_{l},\Phi^{0p}_{c},\Phi^{0p}_{r},\Phi^{1p}_{l},\Phi^{1p}_{c},\Phi^{1p}_{r}\rbrace^{T}$ that is uniquely given by
\begin{align}
\boldsymbol{\Phi} = \Sigma^{-1/2}\boldsymbol{\varphi}
\label{eq:orthWaveSet}
\end{align}
for sufficiently small overlaps $\sum_{i,\alpha}|\Sigma_{i\neq j}^{\alpha\neq\beta}|<|\Sigma_{i=j}^{\alpha=\beta}|=1$.
There are different ways to determine the matrix $\Sigma^{-1/2}$, e.g., using the maximally localized Wannier orbitals~\cite{Marzari2012}, by a Gram-Schmidt orthonormalization procedure~\cite{Barnes2011}, and by symmetric singular value decomposition~\cite{Annavarapu2013}. Here, we used the method of symmetric singular value decomposition which maintains the symmetry of the wavefunctions and imposed the least deformation relative to the original wavefunctions in the least-squares sense. Note that for large inter-dot distances $\boldsymbol{\Phi}\rightarrow\boldsymbol{\varphi}$, thus, both sets of wavefunctions are identical.

With this toolbox of orthonormalized wavefunctions we can now calculate the matrix elements of the Hamiltonian~\eqref{eq:HamZero}-\eqref{eq:HamTun} in the basis given by Eqs.~\eqref{eq:state0app}-\eqref{eq:statesPP} and Eqs.~\eqref{eq:states0}-\eqref{eq:states}
\begin{align}
t_{i\alpha,j\beta} &= \int d^{3}\boldsymbol{r} \Phi^{\alpha*}_{i}(\boldsymbol{r}) \left[\mathcal{K}(\boldsymbol{r})+ \mathcal{V}(\boldsymbol{r})\right] \Phi^{\beta}_{j}(\boldsymbol{r}),\label{eq:MEtun}\\
%C^{\text{in}}_{i\alpha,i\beta} &= \int d^{3}\boldsymbol{r}_{1} \int d^{3}\boldsymbol{r}_{2} \Phi^{\alpha*}_{i}(\boldsymbol{r}_{1})\Phi^{\beta*}_{i}(\boldsymbol{r}_{2})\frac{U_{0}}{|\boldsymbol{r}_{1}-\boldsymbol{r}_{2}|} \Phi^{\beta}_{i}(\boldsymbol{r}_{2})\Phi^{\alpha}_{i}(\boldsymbol{r}_{1}),\label{eq:Coulomb1}\\
%K^{\text{in}}_{i\alpha,i\beta} &=  \int d^{3}\boldsymbol{r}_{1} \int d^{3}\boldsymbol{r}_{2}  \Phi^{\alpha*}_{i}(\boldsymbol{r}_{1})\Phi^{\beta*}_{i}(\boldsymbol{r}_{2})\frac{U_{0}}{|\boldsymbol{r}_{1}-\boldsymbol{r}_{2}|} \Phi^{\beta}_{i}(\boldsymbol{r}_{1})\Phi^{\alpha}_{i}(\boldsymbol{r}_{2}),\label{eq:Coulomb2}\\
C_{i\alpha,j\beta} &=  \int d^{3}\boldsymbol{r}_{1} \int d^{3}\boldsymbol{r}_{2}  \Phi^{\alpha*}_{i}(\boldsymbol{r}_{1})\Phi^{\beta*}_{j}(\boldsymbol{r}_{2})\frac{U_{0}}{|\boldsymbol{r}_{1}-\boldsymbol{r}_{2}|} \Phi^{\beta}_{j}(\boldsymbol{r}_{2})\Phi^{\alpha}_{i}(\boldsymbol{r}_{1}),\label{eq:Coulomb3}\\
K_{i\alpha,j\beta} &=  \int d^{3}\boldsymbol{r}_{1} \int d^{3}\boldsymbol{r}_{2}  \Phi^{\alpha*}_{i}(\boldsymbol{r}_{1})\Phi^{\beta*}_{j}(\boldsymbol{r}_{2})\frac{U_{0}}{|\boldsymbol{r}_{1}-\boldsymbol{r}_{2}|} \Phi^{\beta}_{j}(\boldsymbol{r}_{1})\Phi^{\alpha}_{i}(\boldsymbol{r}_{2}).\label{eq:Coulomb4}
\end{align}
Here, $\mathcal{K}=(-i\hbar\nabla-e\boldsymbol{A})^{2}/2m$ contains the kinetic terms in the presence of a magnetic field $B=\nabla\times\boldsymbol{A}$ using the symmetric gauge $\boldsymbol{A}=B(-y,x,0)^{T}/2$, and $\mathcal{V}$ is the potential given in Eq.~\eqref{eq:pot}.
To compute the Coulomb integrals Eqs.~\eqref{eq:Coulomb3}-\eqref{eq:Coulomb4} we approximated $\frac{U_{0}}{|\boldsymbol{r}_{1}-\boldsymbol{r}_{2}|}\approx\frac{U_{0}}{\sqrt{(x_{1}-x_{2})^{2}+(y_{1}-y_{2})^{2}}}$ in the 2D limit $d\ll R$. These integrals can be evaluated analytically but yield rather unwieldy expressions (not shown here).
%\section{Beyond Hubbard model}
%In this section we show that a full molecular orbital study yields a vanishing dipole moment at the charge noise sweet spot. 

\section{Achievability of the condition $t_{l}=t^{\prime}_{l}$ and $t_{r}=t^{\prime}_{r}$ }

As discussed in the main text, the decoherence properties are best under the condition $|t_{l}|=|t^{\prime}_{l}|$ and $|t_{r}|=|t^{\prime}_{r}|$ since in this case the qubit is also robust against charge noise during the pulse sequences (dynamic sweet spot). Here we show that this condition is realizable in experiments. 

\begin{figure}
\begin{center}
\includegraphics[width=0.6\columnwidth]{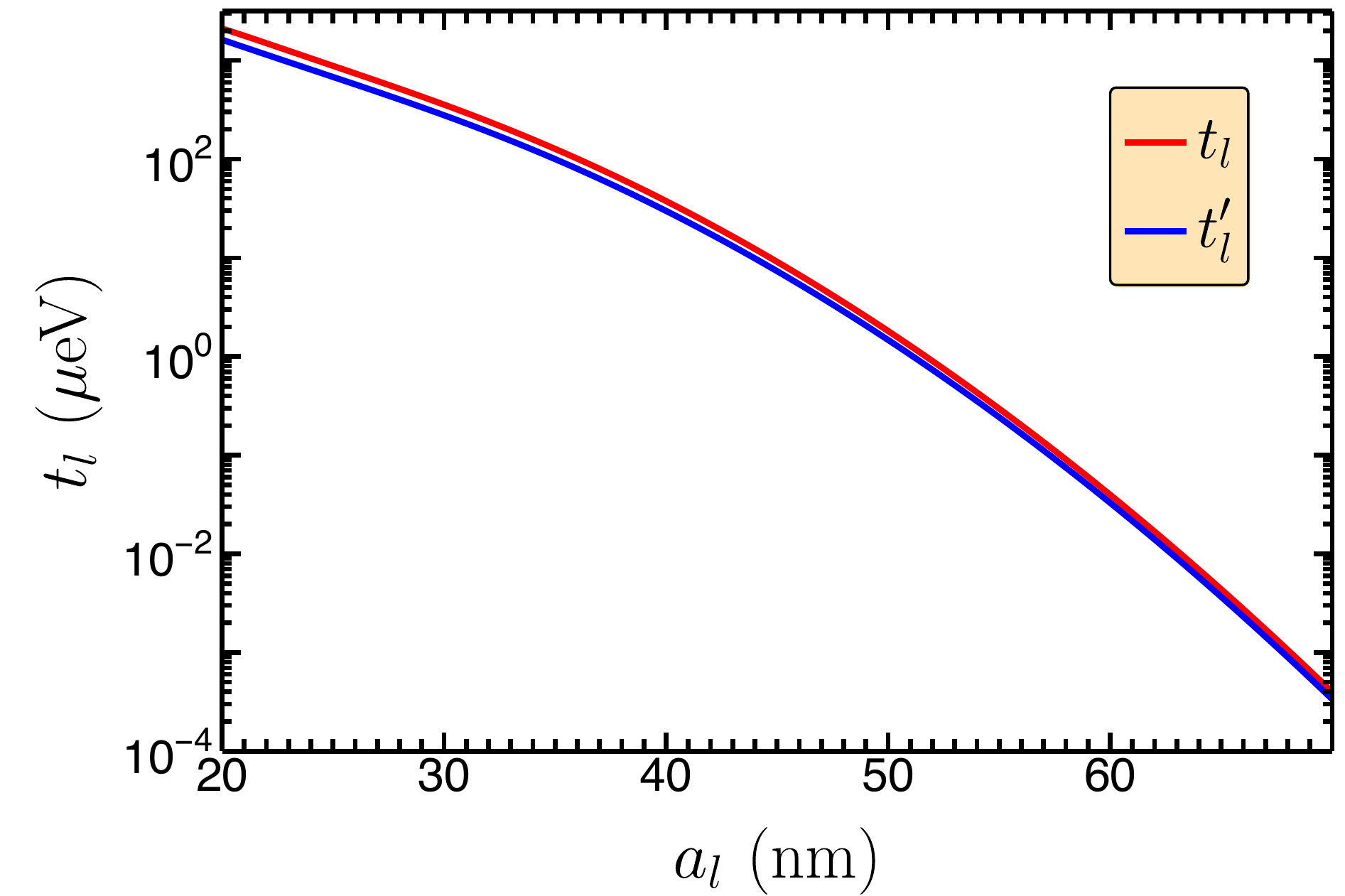}
\caption{Absolute value of the valley-conserving $t_{l}$ (red) and the valley-non-conserving $t^{\prime}_{l}$ (blue) tunnel matrix element between the left and the center dot. At the dip of the blue and red curve the tunneling amplitude is reduced by interference effects. For the simulation the following parameter settings are used; the quantum dot radius $R=\unit[8.17]{nm}$ (corresponds to an orbital energy of $E\st{orb}\approx\unit[6]{meV}$), the magnetic length $l_{B}=\unit[55]{nm}$ (corresponds to a magnetic field $B\approx\unit[210]{mT}$), the two-electron charging energy $E\st{charge}=\unit[3]{meV}$, the valley phases $\phi_{l}^{v}=\unit[0.4]{\pi}$ and $\phi_{r}^{v}=\unit[-0.4]{\pi}$, and valley-orbit parameters [see Eq.~\eqref{eq:VOmatrix}] $ \Delta_{Vs,q}= \Delta_{Vp,q}=\unit[0.2]{meV}$ and $ \Delta_{Oz,q}= \Delta_{O\bar{z},q}=\unit[0.3]{meV}$. Note, that in our simulation the tunnel barrier is set by $a_{l}$ and does not precisely match the inter-dot distance in experimental setups.}
\label{fig:Tun}
\end{center}
\end{figure}

There are two main effects which lead to finite transition amplitudes between a valley ground state and a valley excited state necessary for a finite $t^{\prime}_{l,r}$. The first effect is valley-orbit mixing resulting from imperfect interfaces in the heterostructure. This effect mixes the valley and orbital levels and gives rise to a finite overlap between the electronic wavefunctions. The mixing degree is usually very small, and only a few percent are predicted~\cite{Friesen2010}. Considering only this effect yields only very small valley non-conserving tunneling amplitudes $|t^{\prime}_{l}|\ll |t_{l}|$ and $|t^{\prime}_{r}|\ll |t_{r}|$. The second effect arises from local differences in the orientation of the $z$-valleys due to atomistic steps in the interface. The atomistic interface steps locally change the orientation of the valley pseudo-spin giving rise to a phase difference between valley pseudo-spin of electrons at different lateral position~\cite{Zimmerman2017}. As a result, the valley pseudo-spins in the different quantum dots can have completely different phase factors $\phi^{v}_{q}$ with $q=l,c,r$. Setting $\phi^{v}_{c}=0$, atomistic steps give rise to $|t_{l}|\sim|t^{\prime}_{l}\tan(\phi^{v}_{l}/2)|$ and $|t_{r}|\sim|t^{\prime}_{r}\tan(\phi^{v}_{r}/2)|$ which yields identical tunnel couplings for $\phi^{v}_{l}=\phi^{v}_{r}=\pi/2$. The phases $\phi^{v}_{l}$ and $\phi^{v}_{r}$ are mostly set by the geometry of the interface~\cite{Mi2017}. Tuning can be achieved by changing the position of the electron wavefunctions either by moving the quantum wells by the electrostatic gates or by adjusting the tunnel-barriers.~\cite{Saraiva2009,Zimmerman2017}. Refs.~\cite{Goswami2006,Shi2011,Zhang2013} also show that external magnetic and electric fields provide control over the valley splitting. However, this only weakly affects the phase of the pseudo-spin. A second control mechanism relies on the valley-orbit mixing making use of the fact that the tunneling matrix element between the dots depends on the orbital level. This leads to a difference in the spatial dependence of the tunneling amplitude as seen in Fig.~\ref{fig:Tun}. Here, $|t^{\prime}_{l}|\ll |t_{l}|$ and $|t^{\prime}_{r}|\ll |t_{r}|$ is possible even for $\phi^{v}_{l}\neq \phi^{v}_{r}$.

\section{Electric dipole moment}
 \label{sec:dipole}
 
\begin{figure}
\begin{center}
\includegraphics[width=0.8\columnwidth]{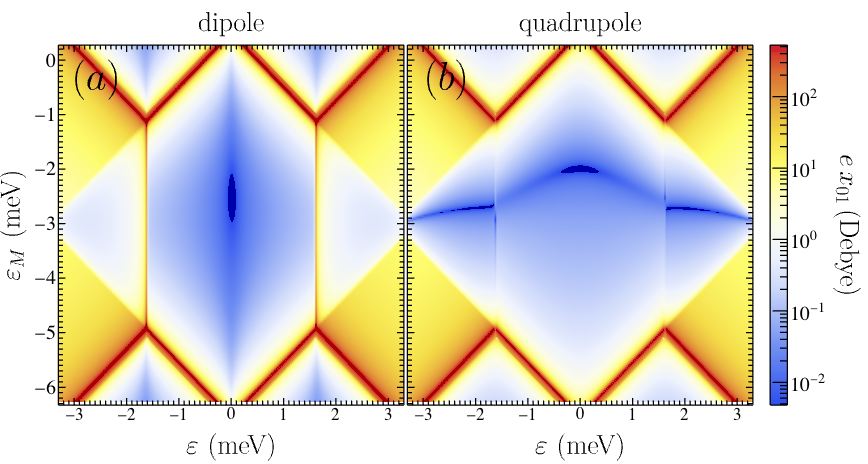}
\caption{Transition matrix element $e\,x_{01}=\bra{0}\hat{x}\ket{1}$ of the position operator~\eqref{eq:dipole} as a function of detuning parameters for $(a)$ a constant electric field $\boldsymbol{E}(\boldsymbol{\hat{x}})=\boldsymbol{E}||\hat{e}_{x}$ and $(b)$ for the electric field $\boldsymbol{E}(\boldsymbol{\hat{x}})=\boldsymbol{E}\,\,\text{sign}(\hat{x})||\hat{e}_{x}$. For the simulation the same parameters are used as in the main text; $t_{l}=t_{r}=\unit[25]{\mu eV}$ and $|t^{\prime}_{l}|=|t^{\prime}_{r}|=\unit[20]{\mu eV}$ and $\arg(t_{l}^{\prime})=\arg(t_{r}^{\prime})=-\pi/2$ which corresponds to the interdot distance $a_{l}=a_{r}=\unit[41.2]{nm}$ in Fig.~\ref{fig:Tun}. }
\label{fig:Dip}
\end{center}
\end{figure}

With the orthonormalized wavefunctions~\eqref{eq:orthWaveSet} we can also calculate the interaction between the qubit and the  electromagnetic field. The interaction Hamiltonian in the dipole approximation reads~\cite{Russ2016}
\begin{align}
H\st{dip}=e\boldsymbol{E}(\boldsymbol{\hat{x}})\cdot\boldsymbol{\hat{x}}
\label{eq:HamDip}
\end{align}
where $\boldsymbol{E}(\boldsymbol{\hat{x}})$ is the position dependent electric field. Following Ref.~\cite{Russ2015b} the dipole transitions are described by the matrix elements of the position operator
\begin{align}
\boldsymbol{\hat{x}}&=\sum_{i,j}\sum_{\alpha,\beta}\sum_{\sigma} \boldsymbol{x}_{i\alpha,j\beta} c^{\dagger}_{i\alpha,\sigma} c_{j\beta,\sigma}.
\label{eq:dipole}
\end{align} 
Analogously to Eq.~\eqref{eq:MEtun}, we find
\begin{align}
\boldsymbol{x}_{i\alpha,j\beta} &= \int d^{3}\boldsymbol{r} \Phi^{\alpha*}_{i}(\boldsymbol{r})\boldsymbol{r}(\boldsymbol{r}) \Phi^{\beta}_{j}(\boldsymbol{r}).
\end{align} 
Fig.~\ref{fig:Dip}~(a) shows the resulting dipole interaction 
\begin{align}
e\,x_{01}=e\bra{0}\hat{x}\ket{1}
\label{eq:transitionElement}
\end{align}
between the two qubit states considering a constant electric field aligned in $x$-direction $\boldsymbol{E}(\boldsymbol{\hat{x}})=\boldsymbol{E}||\hat{e}_{x}$. This corresponds to an architecture where the qubit is connected via the gate $V_{L}$ to the cavity. In Fig.~\ref{fig:Dip}~(b) the results are shown if the qubit is connected via the plunger gate $V_{C}$. In our model this corresponds to $\boldsymbol{E}(\boldsymbol{\hat{x}})=\boldsymbol{E}\,\,\text{sign}(\hat{x})||\hat{e}_{x}$ with $\text{sign}(x)=\frac{x}{|x|}$. 
As expected, there is a small dipole [Fig.~\ref{fig:Dip}~(a)] and quadrupole [Fig.~\ref{fig:Dip}~(b)] moment near the location of the double sweet spot, therefore, protecting the qubit against charge noise. Note, that the dipole and quadrupole moment do not completely vanish due to the choice $|t_{l}|>|t_{l}^{\prime}|$ and $|t_{r}|>|t_{r}^{\prime}|$.
%Clearly both plots show a minimum near the location of the double sweet spot. The deviation from the exact position of the sweet spot arises due to non-vanishing overlaps of the wavefunction~\cite{Russ2015b}.

\section{Qubit decoherence for $t_{l}\neq t_l^{\prime}$ and $t_{r}\neq t_r^{\prime}$}
\label{app:gen}
\begin{figure}
\begin{center}
\includegraphics[width=0.8\columnwidth]{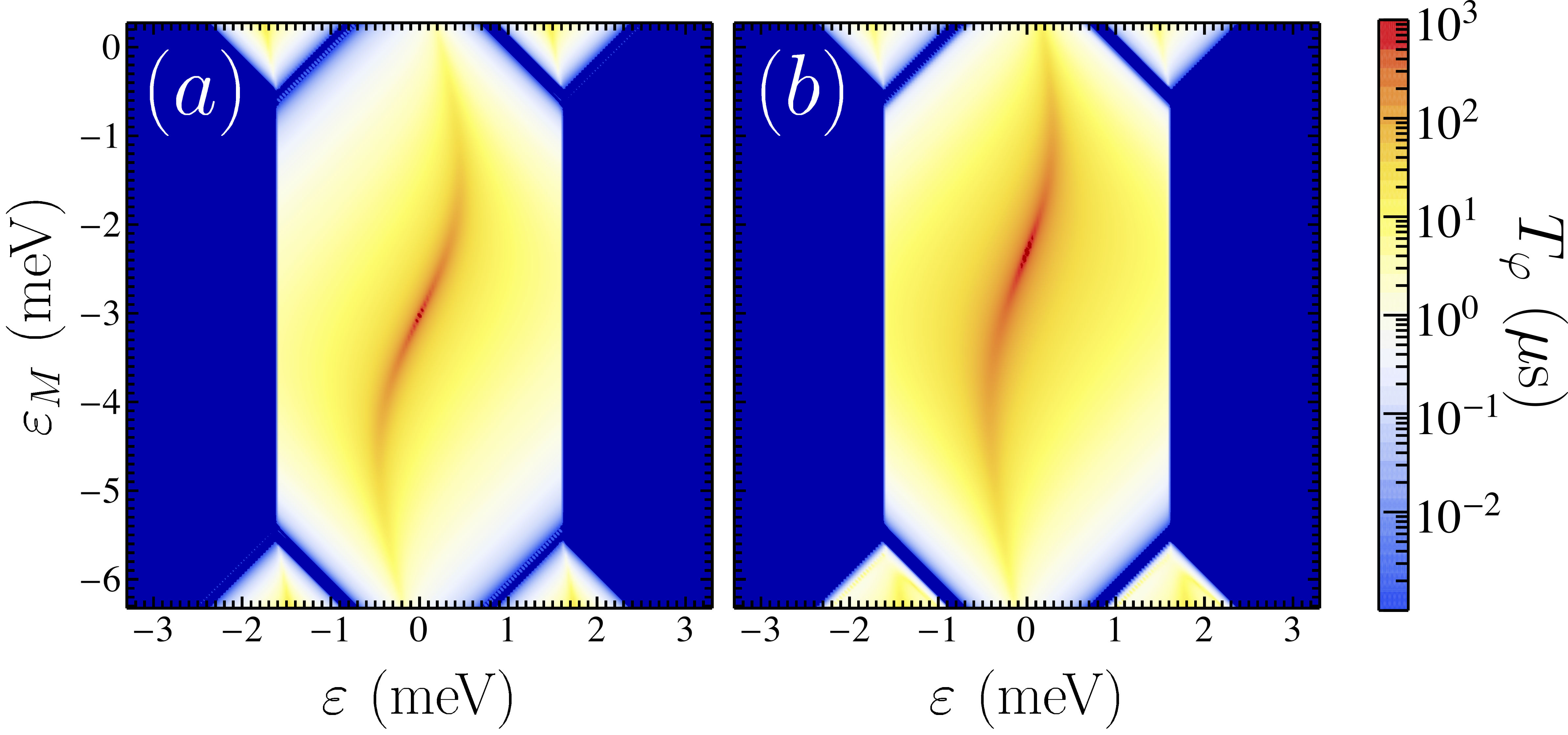}
\caption{Qubit dephasing time $T_{\varphi}$ as a function of $\varepsilon$ and $\varepsilon_{M}$ for (a) $t_{l}=t_l^{\prime}$ and (b) $t_{l}>t_l^{\prime}$. We used the following parameter settings in (a); $t_{l}=t_l^{\prime}=\unit[25]{\mu eV}$, $t_{r}=t_r^{\prime}=\unit[15]{\mu eV}$, $U_{i}=U=\unit[3]{meV}$, $U_{i,j}=U_{C}=\unit[1]{meV}$, $J_{C}=\unit[100]{\mu eV}$, and $E_{V,C}=\unit[150]{\mu eV}$. And in (b) we used the parameter settings; $t_{l}=\unit[22]{\mu eV}$, $t_l^{\prime}=\unit[11]{\mu eV}$, $t_{r}=\unit[15]{\mu eV}$, $t_r^{\prime}=\unit[7.5]{\mu eV}$, $U_{i}=U=\unit[3]{meV}$, $U_{i,j}=U_{C}=\unit[1]{meV}$, $J_{C}=\unit[100]{\mu eV}$, and $E_{V,C}=\unit[150]{\mu eV}$. The simulations indicate the presence of a charge noise double sweet spot with $T_{\varphi}>\unit[100]{\mu s}$ for both parameter settings.}
\label{fig:DephSup}
\end{center}

\end{figure}

In this section we show that the extended sweet spot described in the main text still exist in the general case $t_{l}\neq t_l^{\prime}$ and $t_{r}\neq t_r^{\prime}$. Evidence for the existence of the charge noise double sweet spot is given in Fig.~\ref{fig:DephSup}~(b) which shows a dephasing time $T_{\varphi}$ on the same order as for the location of the double sweet spot defined in the main text for the case $t_{l}=t_l^{\prime}$ and $t_{r}= t_r^{\prime}$ in Fig.~\ref{fig:DephSup}~(a). While this does not proof the existence of a double sweet spot, it does show that one can find working points that are equally protected as in the case $t_{l}=t_l^{\prime}$ and $t_{r}= t_r^{\prime}$. 
%For a numerical proof we also plotted the first derivative of $\omega_{q}$ with respect to $\varepsilon$ and $\varepsilon_{m}$ both showing sweet spots (dark blue lines) which are crossing at some point in the (1,2,1) charge configuration regime. 
The exact location of these sweet spots, however, depends on the tunneling couplings, $t_{l}$, $t_l^{\prime}$, $t_{r}$, and $t_r^{\prime}$ and is given by the conditions
%\begin{align}
%\frac{\partial_{\varepsilon}(J_{1}+E_{V,C}-J_{C}-J_{0})+\partial_{\varepsilon}|J_{x}|}{\omega_{q}^{2}}&\overset{!}{=}0,\\
%\frac{\partial_{\varepsilon_{m}}(J_{1}+E_{V,C}-J_{C}-J_{0})+\partial_{\varepsilon_{m}}|J_{x}|}{\omega_{q}^{2}}&\overset{!}{=}0.
%\end{align}
\begin{align}
\partial_{\varepsilon}\omega_{q}&=0,\\
\partial_{\varepsilon_{M}}\omega_{q}&=0.
\end{align}
These equation cannot be solved analytically in the general case, but we found numerical solutions in the (1,2,1) charge configurations regime for a large set of parameters.

\section{Single qubit operations}
\label{sec:SQO}
In this section we estimate the single-qubit operation time of the QUEX qubit under the effect of periodic driving $V_{C}(t)=V^{0}_{C}+V^{1}_{C}\cos\left(\omega_{D}+\phi\right)$ of the center gate voltage $V_{C}\sim \varepsilon_{M}$. 
Using Eqs.~\eqref{eq:cl}~and~\eqref{eq:cr} one can calculate directly the resulting matrix element between the qubit states from the modulation via Eqs.~\eqref{eq:MEtun}~and~\eqref{eq:Hub}.
Assuming the inter-dot distance $a_{l}=\unit[41.2]{nm}$, a driving voltage $V^{1}_{C}=\unit[1.2]{mV}$, a lever arm $\alpha=\unit[0.1]{meV/mV}$~\cite{Wiel2002}, and $\varepsilon=0$ and $\varepsilon_{M}=\unit[-25]{\mu eV}$ (corresponding to the charge noise double sweet spot), we find the Rabi frequency $A_{\varepsilon_{M}}\approx\unit[0.13]{\mu eV}$. For this choice of parameters, the gate time is estimated to be $\tau_{x}=2\pi\hbar/A_{\varepsilon_{M}}\approx\unit[30]{ns}$ as shown in the main text.
 
 \section{Entangling distant QUEX qubits}
 \label{sec:twoQubit}
 
In this section we estimate the entangling time between two QUEX qubits connected via a shared microwave resonator. The entangling pulse sequence described in the main text is given by~\cite{Srinivasa2016}
 \begin{align}
 U\st{CZ}=R_{a,z}\left(\frac{\pi}{\sqrt{2}}\right)S^{+}_{a}(\pi,\phi)S^{-}_{b}(\frac{\pi}{2},0)S^{-}_{b}(\pi\sqrt{2},\frac{\pi}{2})S^{-}_{b}(\frac{\pi}{2},0)R_{a,z}\left(\frac{\pi}{\sqrt{2}}\right)S^{+}_{a}(\pi,\phi).
 \end{align} 
Here $S_{q}^{\pm}(\phi,\tau)\equiv\exp(-i H_{q,\pm}(g,\phi)\tau/\hbar)$ are ``red'' and ``blue'' sideband transition gates of qubit $q$ generated by the Hamiltonian~\cite{Srinivasa2016,Russ2017}
\begin{align}
H_{\pm} =\frac{g}{2}\left(\E^{\mp i\phi}a^{\dagger} \sigma_{\mp} +\E^{\pm i\phi} a\sigma_{\pm}\right).
\label{eq:2Qpm}
\end{align}
One way to realize this interaction requires driving the qubits with $V^{1}_{C}\cos(\omega_{D}t+\phi)$ at the qubit transition frequency $\omega_{D}=\omega_{q}$ while the qubits are coupled to the resonator via their electric dipole with strength $g_{a}=g_{b}=g$. The interaction Hamiltonian between a single qubit and the cavity in the rotating frame defined by $U\st{r1}=\E^{-i \omega_{D} t(a^{\dagger}a+\sigma_{z}/2) }$ neglecting rapidly oscillating terms is
\begin{align}
H^{\text{rf}}=\Delta_{0}a^{\dagger}a+g\left( \E^{\I\phi} a \sigma_{+}+\E^{-\I\phi} a^{\dagger} \sigma_{-} \right)+\Omega\sigma_{y},
\label{eq:2qRF}
\end{align}
with the effective resonator frequency $\Delta_{0}\equiv\omega_{D}-\omega\st{res}$ and the Rabi frequency of the qubits $\Omega\equiv|gV^{1}_{C}/\Delta_{0}|$. Transforming the Hamiltonian~\eqref{eq:2qRF} into a second rotating frame $U\st{r2}=\E^{-i t(\Delta_{0}a^{\dagger}a+\Omega\sigma_{y}) }$ and subsequently applying the rotation $U\st{rot}=\E^{i\pi\sigma_{x}/4 }$, one obtains~\cite{Srinivasa2016}
\begin{align}
H^{\text{drf}}=& \frac{g}{2} \left[\E^{i\phi}\E^{-i(\Delta_{0}-\Omega)t}\sigma_{+}a + 
\E^{-i\phi}\E^{i(\Delta_{0}-\Omega)t}\sigma_{-}a^{\dagger}+ %\right. \nonumber \\ \left.
\E^{-i\phi}\E^{i(\Delta_{0}+\Omega)t}\sigma_{+}a^{\dagger} +  
\E^{i\phi}\E^{-i(\Delta_{0}+\Omega)t}\sigma_{-}a
\right] \nonumber \\
&+ i \frac{g}{2} \sigma_{z} (\E^{i\phi}\E^{-i\Delta_{0}t}a -  \E^{-i\phi}\E^{i\Delta_{0}t}a^{\dagger})
\label{eq:2qDRF}
\end{align}
Applying the rotating wave approximation for the particular choices $\Omega=\pm\Delta_{0}$ and neglecting rapidly oscillating terms, one obtains the Hamiltonian~\eqref{eq:2Qpm} which enables side-bamd transitions. For this second rotating wave approximation to be valid and therefore minimizing the error, one greatly benefits from the large tunability of the energy gap of the QUEX qubit.

The qubit-resonator coupling is induced by connecting the center gate to the resonator $V_{C}\rightarrow V_{C}(t)=V^{0}_{C}+\alpha V_{0}(a+a^{\dagger})$. Analogously to the procedure for the single-qubit Rabi oscillations (see Sec.~\ref{sec:SQO}), the coupling strength is given by
\begin{align}
g=\alpha \bra{0}H\ket{1}
\end{align}
where $H$ depends on $V_{C}$ via $H_{0}$ and $H\st{tun}$ and where $\alpha$ is the lever arm between the real applied voltage on the center gate and the subsequent changes in the potential~\eqref{eq:pot}, and~\cite{Childress2004}
\begin{align}
V_{0}=\hbar\omega\st{res}\sqrt{\frac{Z_{0}}{\hbar\pi}}
\end{align}
denotes the zero-field fluctuation amplitude of the resonator. 

Considering $\omega\st{res}=\unit[10]{GHz}$ and using realistic parameters~\cite{Samkharadze2016} (extrapolated to $\unit[10]{GHz}$) we find $V_{0}\approx\unit[70.7]{\mu V}$. Assuming $\alpha=\unit[0.1]{meV/mV}$ and $\varepsilon_{M}=\unit[20]{meV}$ allows us to estimate the qubit-resonator coupling strength $g=\unit[2\pi\times 6.4]{MHz}$ and the Rabi frequency (see Sec.~\ref{sec:SQO}) $\Omega=A_{\varepsilon_{M}}=\unit[2\pi\times 110]{MHz}$. The final gate time~\cite{Srinivasa2016} is then $\tau= \left(3+\sqrt{2}\right)\pi/g+2\tau_{z}\approx\unit[339]{ns}$ assuming a single-qubit $z$-rotation gate time $\tau_{z}\approx\unit[5]{ns}$. Note that the gate is performed while operating at the charge noise sweet spot with respect the dipolar detuning $\varepsilon$. Faster gates can be achieved by moving away from the sweet spot.

\clearpage
\section{Higher excitations exchange parameters}
\label{sec:highEx}
In this section we provide all exchange coupling terms for completeness. The explicit expressions are given by
\begin{align}
\tilde{J}_{0,2}&=\frac{1}{2} \left(-\frac{t_l^{\prime} t_{l}^*}{4 \sqrt{2} E_{10}}+\frac{t_l^{\prime} t_{l}^*}{4 \sqrt{2} E_4}+\frac{t_l^{\prime} t_{l}^*}{4 \sqrt{2} (E_{10}-E_{V,C})}+\frac{t_l^{\prime} t_{l}^*}{4 \sqrt{2} (E_{V,C}-E_{4})}\right. \nnb &\left.+\frac{t_{r} t_r^{\prime*}}{4 \sqrt{2} E_6}-\frac{t_{r} t_r^{\prime*}}{4 \sqrt{2} E_{8}}+\frac{t_{r} t_r^{\prime*}}{4 \sqrt{2} (E_{V,C}-E_{6})}-\frac{t_{r} t_r^{\prime*}}{4 \sqrt{2} (E_{V,C}-E_{8})}\right),\\
J_{1,2}&=\frac{1}{2} \left(-\frac{\sqrt{3} |t_l|^2}{8 E_{10}+J_{C}-E_{V,C}}+\frac{\sqrt{3} |t_l^{\prime}|^2}{8 E_{11}+J_{C}-E_{V,C}}+\frac{\sqrt{3} |t_l^{\prime}|^2}{8 E_{4}+J_{C}-E_{V,C}}-\frac{\sqrt{3} |t_l|^2}{8 E_{5}+J_{C}-E_{V,C}}\right. \nnb &\left.-\frac{\sqrt{3} |t_r^{\prime}|^2}{8 E_{6}+J_{C}-E_{V,C}}+\frac{\sqrt{3} |t_r|^2}{8 E_{7}+J_{C}-E_{V,C}}+\frac{\sqrt{3} |t_r|^2}{8 E_{8}+J_{C}-E_{V,C}}-\frac{\sqrt{3} |t_r^{\prime}|^2}{8 E_{9}+J_{C}-E_{V,C}}\right. \nnb &\left.+\frac{\sqrt{3} |t_l|^2}{8 (E_{10}-E_{V,C})}-\frac{\sqrt{3} |t_l^{\prime}|^2}{8 (E_{11}-E_{V,C})}+\frac{\sqrt{3} |t_l^{\prime}|^2}{8 (E_{V,C}-E_{4})}-\frac{\sqrt{3} |t_l|^2}{8 (E_{V,C}-E_{5})}\right. \nnb &\left.-\frac{\sqrt{3} |t_l^{\prime}|^2}{8 (E_{V,C}-E_{6})}+\frac{\sqrt{3} |t_r|^2}{8 (E_{V,C}-E_{7})}+\frac{\sqrt{3} |t_r|^2}{8 (E_{V,C}-E_{8})}-\frac{\sqrt{3} |t_r^{\prime}|^2}{8 (E_{V,C}-E_{9})}\right),\\
J_{2,2}&=\frac{1}{2} \left(-\frac{|t_l|^2}{4 (E_{10}-E_{V,C})}-\frac{|t_l^{\prime}|^2}{4 (E_{11}-E_{V,C})}+\frac{|t_l^{\prime}|^2}{4 (E_{V,C}-E_{4})}+\frac{|t_l|^2}{4 (E_{V,C}-E_{5})}\right. \nnb &\left.+\frac{|t_r^{\prime}|^2}{4 (E_{V,C}-E_{6})}+\frac{|t_r|^2}{4 (E_{V,C}-E_{7})}+\frac{|t_r|^2}{4 (E_{V,C}-E_{8})}+\frac{|t_r^{\prime}|^2}{4 (E_{V,C}-E_{9})}\right),\\
\tilde{J}_{1,3}&=\frac{1}{2} \left(-\frac{\sqrt{\frac{3}{2}} t_l^{\prime} t_{l}^*}{4 E_{11}+J_{C}-E_{V,C}}-\frac{\sqrt{\frac{3}{2}} t_l^{\prime} t_{l}^*}{4 E_{5}+J_{C}-E_{V,C}}+\frac{\sqrt{\frac{3}{2}} t_l^{\prime} t_{l}^*}{4 (E_{11}-2E_{V,C})}-\frac{\sqrt{\frac{3}{2}} t_l^{\prime} t_{l}^*}{4 (2E_{V,C}-E_{5})}\right. \nnb &\left.+\frac{\sqrt{\frac{3}{2}} t_{r} t_r^{\prime*}}{4 E_{7}+J_{C}-E_{V,C}}+\frac{\sqrt{\frac{3}{2}} t_{r} t_r^{\prime*}}{4 E_{9}+J_{C}-E_{V,C}}+\frac{\sqrt{\frac{3}{2}} t_{r} t_r^{\prime*}}{4 (2E_{V,C}-E_{7})}+\frac{\sqrt{\frac{3}{2}} t_{r} t_r^{\prime*}}{4 (2E_{V,C}-E_{9})}\right),\\
\tilde{J}_{2,3}&=\frac{1}{2} \left(\frac{t_l^{\prime} t_{l}^*}{4 \sqrt{2} (E_{11}-E_{V,C})}+\frac{t_l^{\prime} t_{l}^*}{4 \sqrt{2} (E_{11}-2E_{V,C})}+\frac{t_l^{\prime} t_{l}^*}{4 \sqrt{2} (E_{V,C}-E_{5})}+\frac{t_l^{\prime} t_{l}^*}{4 \sqrt{2} (2E_{V,C}-E_{5})}\right. \nnb &\left.+\frac{t_{r} t_r^{\prime*}}{4 \sqrt{2} (E_{V,C}-E_{7})}-\frac{t_{r} t_r^{\prime*}}{4 \sqrt{2} (E_{V,C}-E_{9})}+\frac{t_{r} t_r^{\prime*}}{4 \sqrt{2} (2E_{V,C}-E_{7})}-\frac{t_{r} t_r^{\prime*}}{4 \sqrt{2} (2E_{V,C}-E_{9})}\right),\\
J_{3,3}&=\frac{1}{2} \left(-\frac{|t_l|^2}{2 (E_{11}-2E_{V,C})}+\frac{|t_l^{\prime}|^2}{2 (2E_{V,C}-E_{5})}+\frac{|t_r^{\prime}|^2}{2 (2E_{V,C}-E_{7})}+\frac{|t_r|^2}{2 (2E_{V,C}-E_{9})}\right),
\end{align}
with the energies $E_{i}$ given in Eq.~\eqref{eq:states}.

\bibliography{QUEX}

\end{document}